\documentclass[journal]{IEEEtran}
\IEEEoverridecommandlockouts

\usepackage{amsmath,amssymb,amsfonts}
\usepackage{multicol}
\usepackage{graphicx}
\usepackage{caption}
\usepackage{tablefootnote}
\usepackage{threeparttable}
\usepackage{algorithm}
\usepackage{algpseudocode}
\usepackage{textcomp}
\usepackage{tabularx}
\usepackage{adjustbox}
\usepackage{comment}
\usepackage{xcolor}
\usepackage{float}
\usepackage{caption}
\usepackage{subfigure}
\usepackage{subfloat}
\usepackage{multirow,multicol,booktabs}
\usepackage{algorithm}
\usepackage{algpseudocode}
\graphicspath{ {../figures} }
\usepackage{array}
\usepackage{url}

\usepackage{breakurl}
\usepackage[breaklinks]{hyperref}

\begin{document}


%
\title{
Ecological Robustness-Oriented Grid Network Design for Resilience Against Multiple Hazards
}
\author{Hao Huang\IEEEauthorrefmark{1}, \emph{Member, IEEE,}  \IEEEauthorblockN{Zeyu Mao\IEEEauthorrefmark{1}, \emph{Member, IEEE,} Varuneswara Panyam\IEEEauthorrefmark{2}, Astrid Layton\IEEEauthorrefmark{2}, Katherine R. Davis, \emph{Senior Member, IEEE}\IEEEauthorrefmark{1}}} 




\pagestyle{plain}
\maketitle

\begin{abstract}

Power \textcolor{black}{systems} are critical infrastructure for reliable and secure electric energy delivery. Incidents are increasing, as unexpected multiple hazards ranging from natural disasters to cyberattacks
threaten the security and functionality of 
society. 
Inspired \textcolor{black}{by} resilient ecosystems, 
this paper presents a 
resilient network design approach with an ecological robustness (R\textsubscript{ECO})-oriented optimization 
to improve 
power systems' ability to 
maintain a secure operating state throughout 
unknown hazards.
The 
approach uses 
R\textsubscript{ECO}, a \textit{surprisal}-based metric that captures key features of an ecosystem’s resilient structure,
as an
objective to strategically \textcolor{black}{design} the electrical network.
The approach enables solvability and practicality by introducing
a stochastic-based candidate branch creation algorithm  
and a Taylor series expansion for relaxation of the 
R\textsubscript{ECO} formulation. 
Finally, studies are conducted on the R\textsubscript{ECO}-oriented approach using the IEEE 24 Bus RTS and the ACTIVSg200 systems. Results demonstrate improvement of the system’s reliability under multiple hazards, network properties of robust structure and equally distributed power flows, 
and survivability against cascading failures. From the analysis, we observe that a more redundant network structure with equally distributed power flows benefits its 
resilience.   

\end{abstract}










\section{Introduction}

Power \textcolor{black}{systems} deliver the electric energy that ensures the functionality of modern society.
However, the infrastructure is aging and 
remains 
vulnerable to physical disturbances and natural disasters \cite{shield2021major}, such as the Winter Storm Uri in Texas in 2021.
The integration of communication networks into critical infrastructure enables improved functionality 
but also increases the risk of cyber-originated and combined cyber-physical attacks to
cause unexpected outages~\cite{entso, case2016analysis}. 
The design of a \textit{resilient} grid network is thus an essential foundation for its inherent abilities to withstand such hazards. 
%

Resilience is a property of systems that represents their ability to recover from adverse conditions. 
From a regional transmission operator perspective, Chen \textit{et al.} emphasize the necessity of constructing a robust grid to allow operators to address various contingencies on any given day \cite{chen2020toward}. 
In~\cite{gholami2018toward}, Gholami \textit{et al.} list different areas of resilience enhancement regarding system planning and operations, where \textcolor{black}{long-term planning} 
resilience enhancements lay the foundation for \textcolor{black}{short-term operational} 
resilience enhancements.
Both \cite{panteli2017metrics, panteli2015modeling} find
that redundant and robust network structures are effective for improving power system resilience under extreme conditions. 
These works highlight the importance of network design for enhancing power system resilience, while motivating the need to better understand and characterize the effective use of design against extreme events. Inspired by resilient ecosystems, this work develops a resilience-oriented design approach for large scale power systems that improves their inherently ability to absorb the disturbances from multiple hazards. 
The novelty of this work is to
introduce \textcolor{black}{an optimization based resilient design approach that is realistic, scalable, and extensible} to
translate the long-term resilient trait of ecosystems, ecological robustness
(R\textsubscript{ECO}), into power network design with the consideration of power system constraints, including the
power balance, power flow equations, and operational limits.
The goal of the proposed \textit{R\textsubscript{ECO} Oriented Power Network Design Problem
}
is to strategically add redundancy to power networks and satisfy the constraints of power systems
for improving the system’s resilience.
The main contributions of this paper are as follows:





\begin{itemize}


\item

This paper presents a resilience-oriented approach to improve power systems' inherent ability to tolerate \textit{unexpected} high-impact disturbances and maintain functionality securely. A quantitative resilience metric, R\textsubscript{ECO}, is formulated as an objective for network optimization considering
the power system's constraints
to guide resilient power network design, ahead-of-time without intelligence of the threat. %



 \item
A stochastic-based algorithm to create candidate branches and a Taylor Series Expansion of the logarithm function in the formulation are proposed to scalably solve the optimization, and the R\textsubscript{ECO}-oriented design problem is solved for 24- and 200-bus systems under different scenarios. 
\item
\textcolor{black}{The R\textsubscript{ECO}-oriented power networks are examined under different levels of \textit{N-x} contingencies, 
and their network properties and power flow distributions are analyzed. The analyses show that a more redundant power network structure with more equally distributed power flows contributes to a more resilient power system. }

 \item \textcolor{black}{R\textsubscript{ECO} is shown to be 
 an effective
 metric to help measure and improve
 the inherent resilience of power networks. 
 The 
 formulation
 can guide design of power network structures considering
 power flows
 and ecosystems' resilient traits to achieve power systems' long-term resilience.}

\end{itemize}

Section \ref{review} reviews other resilient power network design approaches and introduces the research objective of this work. Section \ref{sec:formulation} presents related work on unexpected critical \textit{multi-hazards} in power systems and the background of R\textsubscript{ECO}. 
Section \ref{model} introduces the proposed R\textsubscript{ECO}-oriented approach for improving power system resilience through resilient network design. Section \ref{sec:case} applies the R\textsubscript{ECO}-oriented approach to a 24- and a 200-bus 
system, respectively. Section \ref{results} analyzes optimized networks regarding the system reliability under different levels of \textit{N-x} contingencies and network properties. More discussions are in Section \ref{discussion}, and Section \ref{sec:conclusion} concludes the paper.

\section{Background and Research Objectives}
\label{review}
Recently, several works have been proposed to optimally design and plan transmission and distribution systems to improve a system's resilience against natural disasters. 
In \cite{ma2019resilience}, Ma \textit{et al.} present a two-stage stochastic mixed-integer linear program to optimally design the network with minimum investment and minimum expected loss of load during climate hazards. In \cite{lagos2019identifying}, a framework is proposed to analyze the investment in power network enhancements with the evaluation of system resilience under natural disasters. 
In \cite{shahidehpour2021tri}, a tri-level planning approach is proposed to expand and harden the coupled power distribution and transportation systems for improved resilience under random natural disasters with minimum investment. In \cite{garifi2021transmission}, Garifi \textit{et al.} propose a method to harden the power grid structure with minimum investment. The investment decision will improve the grid’s recovery
against natural disasters.
All above works consider the adverse impact of natural disasters with stochastic models and formulate the resilient network design problem from the cost-effectiveness perspective. The improvement of resilience is observed and validated with less loss of load under the adverse scenarios. 
\textcolor{black}{
These works address resilience through different economic incentives for targeted hazards.
However, there is a lack of an accepted and unified \textit{resilience} objective that
captures the inherent property of resilience considering the power network structure. By comparison, the research question addressed 
in this paper
is \textit{how to design a resilient power network structure that can enhance power systems' inherent ability to tolerate disturbances and maintain functionality securely regardless of the source of threats. 
} R\textsubscript{ECO} captures the inherent property of resilience regarding the network design and power flow distribution,
and it represents the inherent ability to absorb disturbances regardless of their sources or causes.}
%
%


\textcolor{black}{As presented in \cite{ton2015more}, \textit{design, preparedness, and planning} have been recognized as the top three needs to enhance grid resilience; 
importantly, design and construction standards for higher performance are required. The research gap addressed in this paper is to integrate the property of resilience into power network design for enhanced inherent resiliency.} 
This paper presents a resilience-driven approach for power network design with its inspiration from naturally resilient ecosystems. The proposed approach translates ecosystems’ survivability and resilience traits to power grids under the guidance of a \textit{quantitative} resilience objective. 


The concept of \textit{resilience} that we adopt dates back to the 1970s when \textit{C.S. Holling} defined the \textit{resilience} in ecology as ``a measure of the ability to absorb changes of variables and parameters in systems \cite{holling1973resilience}.'' Over millions of years' growth and development, ecosystems have survived from various \textit{large-scale} and \textit{unexpected} disturbances, showing the ability to absorb sudden changes in the system and maintain their state. This long-term resilience contributes to an ecosystem's unique network structure, and 
it results in a novel and practical benchmark for robust, sustainable, and resilient human networks design. 
This benchmark is quantified as ecological robustness (R\textsubscript{ECO}) \cite{ulanowicz2009quantifying, Ulan_2004} that adopts a \textit{surprisal} model from \textit{information theory} \cite{gray2011entropy}. By modeling ecosystems as directional graph representations of energy transfer, the optimal R\textsubscript{ECO} recognizes a balance of the pathway \textit{efficiency} and \textit{redundancy} in resilient ecosystems. 
Based on the similarity between ecosystems and power systems, \textcolor{black}{
\cite{panyam2019bioA, huang2020mixed} introduce the potential of
R\textsubscript{ECO} to 
guide power network design for improved reliability.
In \cite{huangreco}, the authors propose a R\textsubscript{ECO}-oriented optimal power flow to improve power systems' survivability against unexpected contingencies. All above works show the potential of applying R\textsubscript{ECO} into power systems to improve resilience. 
However, the approach in \cite{panyam2019bioA} would not be practical to implement. 
For a 14-bus power grid, it is unrealistic to construct 80 branches to improve its resilience. Besides,
~\cite{panyam2019bioA, huang2020mixed} are limited to small-scale power systems due to the mathematical formulation of the R\textsubscript{ECO} and its optimization, while 
\cite{huangreco} only optimizes
the power flow dispatch.
Therefore, this paper introduces a comprehensive R\textsubscript{ECO}-oriented resilient power network design
approach that facilitates scalability and practicality 
for 
large-scale power systems.} 

Three challenges \textcolor{black}{previously impeded the application of R\textsubscript{ECO} for large-scale power network design.}
First, the nature of network design is a mixed-integer problem, which is a typical class of NP hard problem. With the increase of case size, the search domain expands exponentially, which adversely limits the efficiency for finding a global optimal. Second, the optimized networks in \cite{panyam2019bioA} directly connect buses in different voltage levels, which are impractical in power systems. Third, the formulation of R\textsubscript{ECO} involves several layers of logarithm functions which need the input variables to be positive, namely the power flow direction needs to be the same during the solving process. However, power flow direction changes are prevalent in large-scale power systems, and it makes the mixed-integer nonlinear programming (MINLP) problem in \cite{huang2020mixed} invalid for larger power systems. 
To deal with above challenges, this paper first introduces a stochastic based algorithm to create candidate branches with realistic electric parameters for large-scale power systems. It greatly reduces the search domain for the optimal structure and keeps the realism of the network structure. Then, we relax the formulation of R\textsubscript{ECO} with a Taylor series expansion for the logarithm functions. The change of flow direction during the solving process will not cause the problem to be invalid.
\textcolor{black}{This
improves} the solvability and efficiency of the network design problem and ensures the practicality of the optimized resilient network design.

\section{Related Work} 
\label{sec:formulation}

\subsection{Unexpected Multi-Hazard Scenarios}\label{multihazard}

\textit{N-1} reliability is the basic requirement for modern power systems planning and operation \cite{NERCTran}. 
However, the integration of communication networks and the increasing of system size expose power systems to more threats from both cyber and physical domains. Thus, the abruptness of contingencies is increasingly harder to predict \cite{pahwa2014abruptness}. 
In \cite{narimani2021generalized, huang2021toward}, authors have utilized Line Outage Distribution Factors (LODFs) and Group Betweenness Centrality (GBC) to identify sets of critical elements in large scale synthetic grids \cite{syntheticweb}. These sets of critical elements consist of multiple (3 to 8) branches across the \textit{wide} area, which are statistically \textit{unexpected} and can \textit{adversely} disrupt power systems' operation and security. In \cite{sahu2021design}, those \textit{unexpected} \textit{multi-hazards} have been achieved through Man-in-the-Middle attacks (\textit{MiTM}) in a high fidelity cyber-physical power system testbed. Those incidents make the system experience operational stress, threatening grid security and resilience. 
The above \textit{multi-hazard} scenarios provide a touchstone for measuring power system resilience against \textit{unexpected} cyberattacks and natural disasters. Under \textit{unexpected multi-hazards}, 
the system's inherent ability to absorb disturbances can be measured by its resulting
operational violations
as an indicator of its resilience.




\subsection{Background of Ecological Robustness (R\textsubscript{ECO})}



Ulanowicz \textit{et al.} and Fath \textit{et al.} utilize a model of \textit{surprisal} from \textit{Information Theory} \cite{gray2011entropy} to quantify the resilience of ecosystems as R\textsubscript{ECO}. It considers the network structure and the transitions of energy and material among all species over the network \cite{ulanowicz2009quantifying, Ulan_2004,fath2007ecological}. Its formulation represents
a given network's robustness as a function of its energy flow pathway's \textit{redundancy} and \textit{efficiency}. 

\textit{Surprisal} is defined with the following expression, 
\begin{equation}\label{eq:surprisal}
s_i = -k \times log(p_i)	
\end{equation}
where $s_i$ is one's ``surprisal'' at observing an event \textit{i} that occurs with probability $p_i$, and $k$ is a positive scalar constant \cite{shannon2001mathematical}. 
%

The \textit{indeterminacy} ($h_i$) of an event $i$ is then formulated as the product of the presence of an event $p_i$ and its absence $s_i$:
\begin{equation}
\label{intermittency}
h_i = -k \times p_i \times log(p_i)	= s_i \times p_i 
\end{equation}

It measures how likely the event $i$ will change for a given event $i$, if we know the probability of event $i$ will occur ($p_i >> 0$) and the surprisal of event $i$ that the system is doing something else most of the time ($s_i >> 0$). 
It can be interpreted as follows: for a given system, those low probability events can cause high impacts to the system, because they happen so rarely that the system doesn't expect; high probability events possess a low impact because they occur often and the system adapts to them \cite{bodini2002towards}.


With the above models of \textit{surprisal} and \textit{indeterminacy}, R\textsubscript{ECO} is formulated with the following metrics.

\begin{figure}[h!]
\centering
\includegraphics[trim={1.2mm 1.2mm 1.2mm 1.2mm}, clip,width=0.98\linewidth]{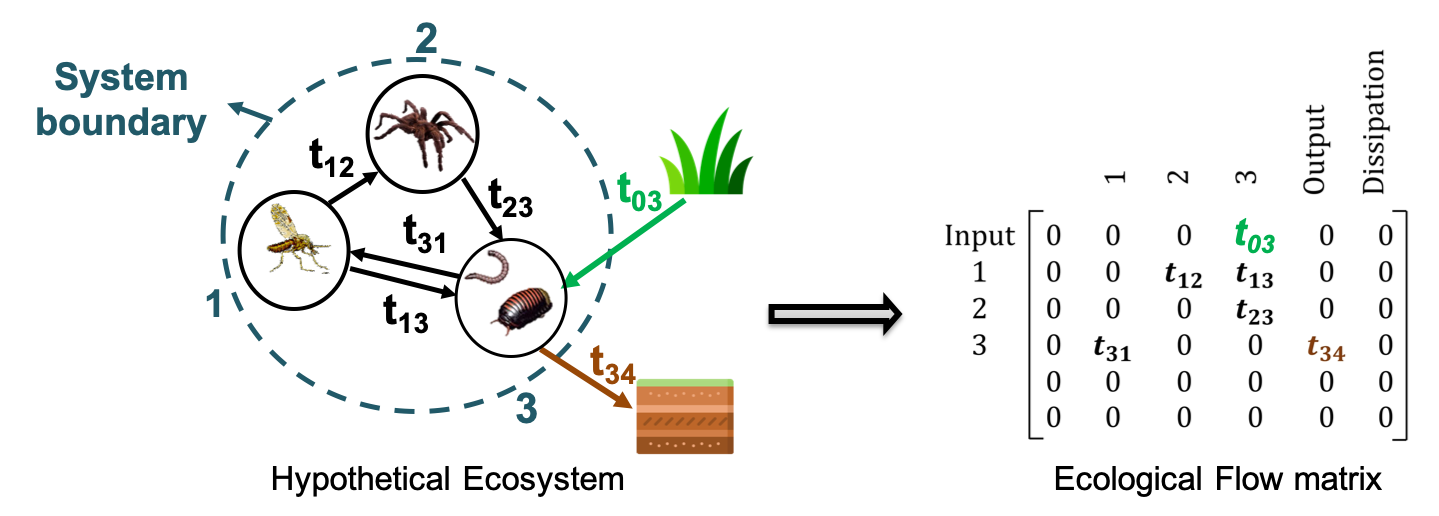}
  \caption{\textcolor{black}{The conversion of a hypothetical ecosystem into Ecological Flow Matrix. Replicated from \cite{panyam2019bioA}.}} 
  \label{hypotheticEFM}
  
\end{figure}

The \textbf{Total System Throughput} (\textit{TSTp}) is the sum of all flows within the system, which represents the system size \cite{ulanowicz2012growth}, 

\begin{equation}
\label{eq:tstp}
\textit{TSTp} = \sum_{i=1}^{N+3}\sum_{j=1}^{N+3}T_{ij}
\end{equation}

\noindent where \textit{T\textsubscript{ij}} is the entry in \textit{Ecological Flow Matrix} (\textit{EFM}) [\textbf{T}]. \textcolor{black}{Following the ecologists' modeling of food webs, the \textit{EFM} is constructed with 
a 
system boundary. Fig \ref{hypotheticEFM} shows a hypothetical ecosystem and its conversion to [\textbf{T}]. The actors (species) that exchange energy based on a \textit{prey-predator} relationship are within the system boundary, and the energy providers, energy consumers, and energy dissipation are placed outside of the system boundary \cite{ulanowicz2012growth}.
Thus, [\textbf{T}] is a square ($N$+3) $\times$ ($N$+3) matrix containing flow magnitudes of transferred energy. \textit{N} is the number of actors inside the system boundary, and the extra three rows/columns represent the system inputs, useful system exports, and dissipation or system exports \cite{Layton2014}. It captures the energy interactions within and 
across the system boundary.} 

The \textbf{Ascendency} (\textit{ASC}) measures the scaled mutual constraint for system size and flow organization that describes the process of ecosystems' growth and development \cite{ulanowicz1980hypothesis} with following expression, 

\begin{equation} \label{eq:asc}
\textit{ASC} = -\textit{TSTp} \times \sum_{i=1}^{N+3} \sum_{j=1}^{N+3} \Bigg( \frac{T_{ij}}{\textit{TSTp}} log_2 \Bigg( \displaystyle \frac{T_{ij} \textit{TSTp}} {T_{i} T_{j}} \Bigg) \Bigg)
\end{equation}


\noindent \textcolor{black}{where $\frac{T_{ij}}{TSTp}$ is recognized as the probability of an event that interrupts ${T_{ij}}$ with respect to all flows circulated in the system. 
$\frac{T_{ij}TSTp}{T_{i}T_{j}}$ measures the conditional probability of joint event \textit{i} and \textit{j} with knowledge of source node (\textit{i}) and end node (\textit{j}), where $T_{i}=\sum_{m=1}^{m=N+3}T_{im}$ and $T_{j}=\sum_{n=1}^{n=N+3}T_{nj}$. With the model of 
\textit{indeterminacy} (Eqn. \ref{intermittency}), the sum of $\frac{T_{ij}}{TSTp}log_{2}(\frac{T_{ij}TSTp}{T_{i}T_{j}})$ multiplies with \textit{TSTp}, giving a dimensional version of network uncertainty.} For the same size systems, a higher value of \textit{ASC} means that a network has fewer options of pathways for flows moving from any one actor to another, resulting 
in a lower level of uncertainty.

The \textbf{Development Capacity} (\textit{DC}) is the upper bound of \textit{ASC} as the development and growth of ecosystems are limited \cite{ulanowlcz1990symmetrical}, 
\begin{equation} \label{eq:dc}
\textit{DC} = -\textit{TSTp} \sum_{i=1}^{N+3} \sum_{j=1}^{N+3} \Bigg(\displaystyle \frac{T_{ij}}{\textit{TSTp}} log_2 \Big(\displaystyle \frac{T_{ij}}{\textit{TSTp}}\Big)\Bigg)
\end{equation}

%

Similar to \textit{ASC}, 
\textit{DC} is also the aggregate uncertainty, but without considering the source and end nodes. 
It captures the aggregated impacts (uncertainty) from all events (surprisals).

Then, R\textsubscript{ECO} is then formulated as follows:

\begin{equation}\label{eq:r}
    \textit{R\textsubscript{ECO}} = -\bigg( \displaystyle \frac{\textit{ASC}}{\textit{DC}} \bigg) ln\bigg(\displaystyle \frac{\textit{ASC}}{\textit{DC}}\bigg)
\end{equation}

The ratio of \textit{ASC} and \textit{DC} reflects the \textit{pathway efficiency} for a given network, while its natural logarithm shows the network’s \textit{pathway redundancy} \cite{ulanowicz2009quantifying}. Thus, R\textsubscript{ECO} is a function of these two opposing but complementary attributes,
where their balance 
achieves the optimal R\textsubscript{ECO} 
that directly affects a system's long-term survival~\cite{ulanowicz2009quantifying}. \textcolor{black}{ Multi-element contingency analyses in systems controlled for optimal R\textsubscript{ECO}
\cite{huangreco} 
have shown the ability for R\textsubscript{ECO} to account for the presence of unknown events, or interruptions, that can happen in the system.}

\section{Ecological Robustness-Oriented Approach for Resilient Power Networks}\label{model}

Modeling a power system analogous to 
an ecosystem enables construction of [\textbf{T}] with real power flows which enables R\textsubscript{ECO} optimization and analysis~\cite{panyam2019bioA, huang2020mixed}. 
The analogy adopted between power grids and food webs sets the food web \textit{actors} as 
generators and buses, the \textit{system inputs} as energy supplied to generators from outside the system boundaries, the \textit{useful exports} as loads (demand), and the \textit{dissipation} as real power losses. Fig \ref{5case} shows an exemplar [\textbf{T}] for a grid with \textit{n} generators and \textit{m} buses. 


With [\textbf{T}] constructed from real power flows, \textit{DC} 
estimates the aggregated impacts of all events as the maximum power flow changes that can happen in the system. \textit{ASC} estimates the dependence between events, and
R\textsubscript{ECO} estimates the robustness 
of the system.
%
Then, 
by including R\textsubscript{ECO} 
as an objective to guide network design,
the optimized networks can better inherently absorb disturbances while maintaining functionality securely,
thus improving their resilience.

\subsection{Mixed-Integer Optimization Model}
\label{optimizationmodel}
The \textit{R\textsubscript{ECO}-Oriented Power Network Design Problem} is built upon the Transmission Network Expansion Planning (TNEP) problem and implemented using \texttt{PowerModels.jl} with the objective of achieving optimal \textit{R\textsubscript{ECO}}. The problem is formulated through Equation (\ref{7a})-(\ref{7k}) with the direct current (DC) power flow model. 
\textcolor{black}{The novelty of this model is 
integrating knowledge of this resilient property from ecosystems with the physics in power systems for resilient power network design.}

\textbf{Objective}: 

\begin{equation}
Max(\textit{R\textsubscript{ECO}})
\label{7a}
\end{equation}

\textbf{Subject to}:

\textcolor{black}{
\begin{equation}
\begin{split}
[\textbf{T}]=f(P_{ij},P_{gen_{i}},P_{load_{i}}, 
\alpha_{ij}) \;\;\;\;\;\;\;\;\; \\ =
\begin{bmatrix}
0, & P_{gen_{i}},&0, &... & ...& 0\\
0, & ...&P_{gen_{i}},&0,&...  & 0\\
 0, & ...& ... & ...& ... & 0\\
0, & ...&P_{ij},& ... &P_{load_{i}}, & 0\\
 0, & ...& ... & ...& ... & 0\\
0, & ... &... & \alpha_{ij}P_{ij}, &P_{load_{i}}, & 0\\
 0, & ...& ... & ...& ... & 0
\end{bmatrix}
\label{7n}
\end{split}
\end{equation}}

\begin{equation}\label{eq:r1}
    \textit{R\textsubscript{ECO}} = -\bigg( \displaystyle \frac{\textit{ASC}}{\textit{DC}} \bigg) ln\bigg(\displaystyle \frac{\textit{ASC}}{\textit{DC}}\bigg)
\end{equation}

\begin{equation} \label{eq:asc1}
\textit{ASC} = -\textit{TSTp} \sum_{i=1}^{N+3} \sum_{j=1}^{N+3} \Bigg( \frac{T_{ij}}{\textit{TSTp}} log_2 \Bigg( \displaystyle \frac{T_{ij} \textit{TSTp}} {T_{i} T_{j}} \Bigg) \Bigg)
\end{equation}

\begin{equation} \label{eq:dc1}
\textit{DC} = -\textit{TSTp} \sum_{i=1}^{N+3} \sum_{j=1}^{N+3} \Bigg(\displaystyle \frac{T_{ij}}{\textit{TSTp}} log_2 \Big(\displaystyle \frac{T_{ij}}{\textit{TSTp}}\Big)\Bigg)
\end{equation}

\begin{equation}
\label{eq:tstp1}
\textit{TSTp} = \sum_{i=1}^{N+3}\sum_{j=1}^{N+3}T_{ij}
\end{equation}

\begin{equation}
P^{l}_{ij} \leqslant P_{ij} \leqslant P^{u}_{ij} \; (\forall (i,j) \in \mathcal{B} \cup \mathcal{NB}) 
\label{7c}
\end{equation}
\vspace{-0.8cm}

\begin{equation}
P^{l}_{gen_{i}} \leqslant P_{gen_{i}} \leqslant P^{u}_{gen_{i}}  \; (\forall i \in \mathcal{G})
\label{7d}
\end{equation}
\vspace{-0.8cm}

\begin{equation}
P_{ij} =B_{ij}(\theta_i-\theta_j) \; (\forall (i,j) \in \mathcal{B} )
\label{dcflow}
\end{equation}

\begin{equation}
P_{ij} =\alpha_{ij}B_{ij}(\theta_i-\theta_j) \; (\forall (i,j) \in \mathcal{NB} )
\label{dcflow_candidate}
\end{equation}

\begin{equation}
P_{i} =P_{load_{i}}-P_{gen_{i}}= \sum_{j}P_{ij} \;  (\forall j \in \mathcal{M})
\label{7k}
\end{equation}





\noindent where $\mathcal{B}$ is the set of existing branches,
$\mathcal{NB}$ is the set of candidates of new branches,
$\mathcal{M}$ is the set of buses,
and $\mathcal{G}$ is the set of generators; $P^{l}_{ij}$ and $P^{u}_{ij}$ are the lower and upper bound of branch limit, respectively; $P^{l}_{gen_{i}}$ and $P^{u}_{gen_{i}}$ are the lower and upper bound of generator output, respectively.

\begin{figure}[h!]
\centering
\includegraphics[trim={1.2mm 1.2mm 1.2mm 1.2mm}, clip,width=0.95\linewidth]{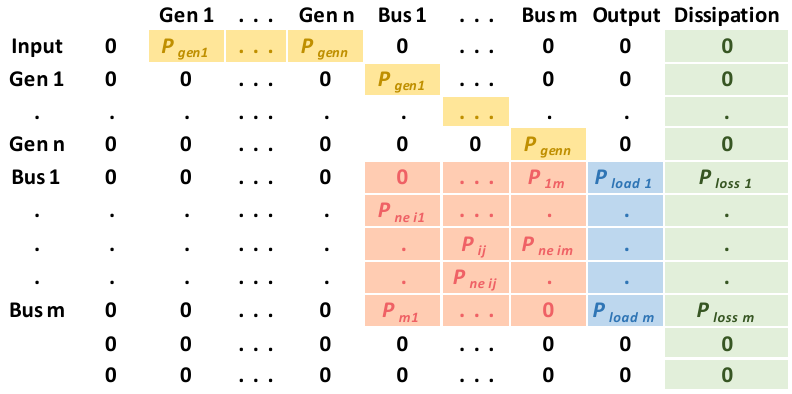}
  \caption{\textcolor{black}{An exemplar of \textit{Ecological Flow Matrix} [\textbf{T}] for a grid with \textit{n} generators and \textit{m} buses. The entries of [\textbf{T}] are $P_{gen_{i}}$, $P_{ij}$ and $P_{ne_{ij}}$, $P_{load_{i}}$, $P_{loss_{i}}$. The $P_{gen_{i}}$ is the real power output from generator $i$, which locates at the input row and the flow between generator and corresponding bus. The generators are treated as lossless with no 
  dissipation.
  The $P_{load_{i}}$ and $P_{loss_{i}}$ are the real power consumption and real power loss, respectively, at Bus $i$. The $P_{ij}$ and $P_{ne_{ij}}$ are the real power flows on the corresponding existing branch and candidate branch, respectively. Entries with zero values mean there is no power flow interaction among buses and generators.}}
  \label{5case}

  
\end{figure}


The TNEP problem is formulated as a mixed-integer optimization problem where each candidate branch has a binary decision variable $\alpha$\textsubscript{\textit{ij}} for the candidate branch from bus $i$ to bus $j$. The initial value of $\alpha$\textsubscript{\textit{ij}} equals to zero if the corresponding branch does not exist in the original network. 
If $\alpha$\textsubscript{\textit{ij}} after optimization equals one, the branch is built to reach a maximum R\textsubscript{ECO}.

\textcolor{black}{The calculation of R\textsubscript{ECO} depends on [\textbf{T}] as expressed in Eqn. \ref{7n} with generator real power output $P_{gen_{i}}$ of each generator, real power flow $P_{ij}$ and $P_{ne_{ij}}$ from existing branches and candidate branches, power consumption at each load $P_{load_{i}}$, 
and binary decision variables $\alpha_{ij}$ for candidate branches. Fig. \ref{5case} illustrates the detailed formulation of [\textbf{T}] using the above variables. Power flow dispatch ($P_{ij}$ and $P_{ne_{ij}}$) depends on the real ($P_{i}$) and reactive power ($Q_{i}$) injection at each bus, bus voltage (voltage magnitude $V_{i}$, voltage angle $\theta_{i}$), and the network structure ($\alpha_{ij}$) \cite{DuncanGlover2012}. 
In this formulation, a DC power flow model is used, so 
voltage magnitude $V_{i}$ is one, reactive power $Q_{i}$ and real power losses $P_{loss_{i}}$ are zero. 
Therefore, the decision variables of the \textit{R\textsubscript{ECO}-Oriented Power Network Design Problem} include each generator's real power input $P_{gen_{i}}$, voltage angle $\theta_{i}$ of each bus, and the binary decision variable $\alpha_{ij}$ for candidate branches. In this way, the proposed R\textsubscript{ECO}-oriented approach will optimize the network structure ($\alpha_{ij}$) and power flow dispatch ($P_{gen_{i}}$ and $\theta_{i}$) to maximize R\textsubscript{ECO}.
Eqn. (\ref{eq:r1})-(\ref{eq:tstp1}) are the calculation of R\textsubscript{ECO} using [\textbf{T}] through several layers of logarithm functions. Eqn. (\ref{7c})-(\ref{7k}) are the power flow constraints for operating limits and power balance. This MINLP problem is thus a nonlinear non-convex optimization problem.}

\subsection{Stochastic Based Candidate Branches Creation}
\label{Algorithm1}
\textcolor{black}{
In the proposed MINLP problem in Section \ref{optimizationmodel}, rather than considering all potential branches to build,
a set of candidate branches is considered.
This represents 
that in reality, the planners have some \textit{a priori} information about new lines to consider.
To represent the impact of the variability of such a set on the formulation and its solution, assuming here that we do not know and cannot control what lines they will choose, and to serve as a proxy for this set, 
we implement the hypothetical 
scenario where this set is chosen randomly.  
%
This selection mechanism can be considered as a worst case scenario, which is suitable to study, as the true planners may be able to choose a better set than a random selection. 
Hence, by demonstrating algorithm's effectiveness even when it is assumed that no information is given about the candidate branch locations, it shows the potential of the approach to be this good or better in reality. This introduces opportunity for future study.
}
\textcolor{black}{Since
the test cases do not include a candidate branch set, 
Algorithm \ref{create} is used,
with suitable electric parameters for each branch based on the existing grid information. 
This fills the gap of the lacked information for candidate branches to expand power networks. 
Unlike \cite{panyam2019bioA, mehrtash2021new} that use
heuristics or pre-screening 
methods 
to find
an
optimal network structure,
Algorithm \ref{create} is a stochastic approach to select candidate lines that supports direct inclusion of
R\textsubscript{ECO} with power system constraints to 
optimize power network structure for inherent resilience.}


\textcolor{black}{The input for Algorithm \ref{create} is the bus and branch information of a given power network, including identifier information, voltage levels, and branches' electric parameters. Algorithm \ref{create} first classifies the existing branches into different voltage levels. 
The normal distribution is then 
used to represent the real-valued random variables. 
Thus, we take branches' electric parameters, including series resistance ($R$), series reactance ($X$), and shunt conductance ($C$), and capacity (MVA limit), as real-valued random variables following the \textit{normal distribution}.
Based on the case information, Algorithm \ref{create} generates \textit{normal distributions} for different electric parameters of branches at each voltage level. 
Algorithm \ref{create} takes a 40\% confidence interval to create valid and different electric parameters of $R$, $X$, and $C$ in per unit for candidate branches in our case studies. The candidate branches' capacity are twice the average capacity of existing cases' branches. 
From ecologists' perspective, power networks are more efficient than redundant. Each network 
has a corresponding value of R\textsubscript{ECO}, and any new branch could contribute to the improvement of R\textsubscript{ECO}. %
In selecting the initial candidate branches, we hypothesize that all network structures have \textit{approximately} the same probability of being the most resilient network based on R\textsubscript{ECO}; hence, all branches are assumed to have the same probability to be selected 
using the \textit{uniform distribution}. 
Algorithm \ref{create} will select \textit{M} candidate branches from all possible branches with the \textit{uniform distribution} to reduce the searching domain. With a specified number of candidate branches (\textit{M}), the probability of selecting candidate branches is ($\frac{1}{M}$)}


Additional information such as geographic location, cost, and government policies can further improve the realism for choosing candidate branches and validating the cost-effectiveness for the network construction. \textcolor{black}{Such information can help stakeholders determine the candidate branches instead of using Algorithm 1. The material, electrical parameters, and construction cost of candidate branches can also then be practically and accurately estimated.} 

One potential issue that may arise 
when adding 
branches 
is the so-called Braess paradox where adding one or more roads can cause congestion and slow down the traffic \cite{steinberg1983prevalence}. A similar situation has been observed in power systems where added branches induced congestion in the system \cite{blumsack2006braess, schafer2022understanding}.  %
The Braess paradox is avoided in the proposed R\textsubscript{ECO}-oriented power network design in Section \ref{optimizationmodel} because the optimization model will reject the branches that can cause congestion in the system.
The results and analyses from the case studies also show this.



\begin{algorithm}[t!]
\begin{algorithmic}
\caption{Stochastic Based Realistic Candidate Branches Selection and Creation}
\label{create}

\State Input = All branches' information from the case, the total number of candidate branches ($M$)
\State Classify branches based on the voltage level
\While {The number of candidate branches $<$ $M$ }
\For {Each Voltage Level}
\State Collect the branch information for all parameters
\For {Each Parameter}
\State Compute the \textit{mean} ($\mu$) and \textit{variance} ($\sigma^{2}$)
\State Generate a \textit{Normal Distribution} ($\mathcal{N}$($\mu$,\,$\sigma^{2}$)\,) 
\EndFor 
\EndFor 
\State Select the \textit{from bus} and \textit{to bus} at the same voltage level using a \textit{Uniform Distribution} ($\mathcal{U}$(0, M)) 
\State Insert the parameter for the candidate branch from the \textit{Normal Distribution} ($\mathcal{N}$($\mu$,\,$\sigma^{2}$)\,).

\EndWhile

\end{algorithmic}
\end{algorithm}

\subsection{Relaxation of the Ecological Robustness Formulation}



\textcolor{black}{The formulation of R\textsubscript{ECO} involves with several layers of logarithm functions, whose hard constraint is that its inputs must remain
positive
during the solving process for the proposed optimization problem using state-of-art MINLP solvers. However, the inputs for calculating R\textsubscript{ECO} are the power flows, and their directions can be reversed during the solving process. 
In \cite{huang2020mixed}, its
formulation fails to capture the feasible space for even small scale power systems, since the inputs are not constantly positive for the logarithm functions during the solving process. 
This creates a problem for large cases, where flow direction changes are more prevalent during the solving process.} 
A Taylor Series Expansion of the natural logarithm function is thus used here to relax the formulation of R\textsubscript{ECO} to ensure the feasibility of the proposed R\textsubscript{ECO}-oriented power network design problem. 




Considering the domain for the expansion, this paper utilizes the following relaxation, with \textit{x} $>$ 0 \cite{jin2007mathematical}: 

\begin{equation}
\begin{split}
ln(x) & =2\sum_{n=1}^{\infty}\frac{((x-1)/(x+1))^{(2n-1)}}{(2n-1)} \\
& =2[\frac{(x-1)}{(x+1)}+\frac{1}{3}(\frac{(x-1)}{(x+1)})^3+\frac{1}{5}(\frac{(x-1)}{(x+1)})^5+... ]
\label{relax1}
\end{split}
\end{equation}

The logarithm function has a base of 2 in Equations (\ref{eq:dc}) and (\ref{eq:asc}). Using a property of logarithm functions,
\begin{equation}
log_2 (x) =\frac{ln(x)}{ln(2)}
\label{changebase}
\end{equation}

\noindent the Taylor Series Expansion of $log_2(x)$ can be expanded:
\begin{equation}
log_2 (x) =\frac{2}{ln(2)}[\frac{(x-1)}{(x+1)}+\frac{1}{3}(\frac{(x-1)}{(x+1)})^3+\frac{1}{5}(\frac{(x-1)}{(x+1)})^5+... ]
\label{relax2}
\end{equation}

By adapting the first order Taylor Series Expansion of Equations (\ref{relax1}) and (\ref{relax2}) into Equation (\ref{eq:r1}) - (\ref{eq:dc1}), the formulation of R\textsubscript{ECO} can keep valid even with flow direction changes during the optimization process. The above formulation requires the input \textit{x} not equal to -1. The \textit{x} for the logarithm function in Equations (\ref{eq:r1})-(\ref{eq:dc1}) are $\frac{ASC}{DC}$, $\frac{T_{ij}TSTp}{T_{i}T_{j}}$, and $\frac{T_{ij}}{TSTp}$, respectively. 
\textcolor{black}{$\frac{ASC}{DC}$ and $\frac{T_{ij}}{TSTp}$ are guaranteed within (-1,1) and $\frac{T_{ij}TSTp}{T_{i}T_{j}}$ is not equal to -1 for power systems.}
Then, the relaxed R\textsubscript{ECO} in the proposed approach can thus be solved with large power grid networks. 


%
\section{Case Studies}\label{sec:case}

This section applies the R\textsubscript{ECO}-oriented approach for two power system cases: the IEEE 24 Bus RTS \cite{zimmerman2016matpower} and the 200-bus synthetic grid from \cite{syntheticweb}, 
 to improve their inherent ability to tolerate disturbances and maintain functionality securely. 
Algorithm \ref{create} created 50, 100, 150, and 200 candidate branches for each case, respectively. \textcolor{black}{Each case has a unique set of candidate branches, and each set of candidate branches does not belong to the others. For example, the set with 100 candidate branches does not include the set with 50 candidate branches.} These candidate branches constitute $2^{50}$, $2^{100}$, $2^{150}$, and $2^{200}$ different network structures to find the optimal R\textsubscript{ECO}-oriented structure through solving the proposed R\textsubscript{ECO}-oriented design problem. 


The candidate branches are selected with the highest voltage rating for each case, since the highest voltage transmission lines are the backbone of the system for power transfer. The proposed approach (\textcolor{black}{Equation \ref{7a}-\ref{7k}}) not only solves the network structure (\textcolor{black}{$\alpha_{ij}$}), it also solves the optimal power flow dispatch with an output vector of generator real power and bus voltage setpoints (\textcolor{black}{$P_{gen_{i}}$ and $\theta_{i}$}). The resultant network design is analyzed with both the optimized network structure 
and the optimized network structure with the output vector, \textcolor{black}{respectively.
Thus, there two \textit{types} of optimized networks analyzed for each scenario under each case study. The naming convention used for each network follows the pattern of \textit{Original Case Name}-\textit{Number of Candidate Branches}-\textit{Structure/Str-OPF}. For the \textit{-Structure} cases, they are the optimized network structure with the selected branches ($\alpha_{ij}$) from the solution to analyze the optimized resilient network structure under \textit{original operating points}; while for the \textit{-Str-OPF} cases, they are the optimized network structure with the operating points of each generator's output and bus voltage ($\alpha_{ij}$, $P_{gen_{i}}$ and $\theta_{i}$).}
The detailed case information have been made publicly available at \cite{BioCases}. 

The solver for the MINLP problem uses \textit{Ipopt}~\cite{wachter2006implementation}, \textit{Juniper}~\cite{kroger2018juniper}, and \textit{Cbc}~\cite{cbc}. Since the MINLP in Section \ref{optimizationmodel} is a nonlinear non-convex problem, the solver can only find the local optimal point. 
All the problems were solved using a laptop with a 2.4 GHz processor and 8 GB memory. The value of the \textbf{Optimal R\textsubscript{ECO}} from the solver is 0.3431. It is the mathematical optimal value of R\textsubscript{ECO} with the Taylor Series Expansion. The results in Table \ref{24 bus result} and \ref{200 result} show the \textbf{Achieved R\textsubscript{ECO}}, the \textbf{Operational Cost}, the \textbf{Number of Added Branches}, the \textbf{Real Power Losses}, the \textbf{Reactive Power Losses}, and the \textbf{Computation Time} for the IEEE 24 Bus RTS system and ACTIVSg200 system, respectively. The \textbf{Achieved R\textsubscript{ECO}} is based on the optimized network structure with/without the output vector of generator real power output and bus voltage after solving the power flows of optimized case with the alternative current (AC) power flow model. 
\textcolor{black}{The \textbf{Operational Cost} is based on the marginal cost ($C_{i}$) \$/MWhr and generator's output ($P_{i}$) MW with Eq. (\ref{cost}), so the unit is \$/hr.} 

\begin{equation}
Cost=\sum_{i=1}^{\mathcal{G}} C_{i}(P_{i}) 
\label{cost}
\end{equation}

\subsection{IEEE 24 Bus Reliability Test System (RTS)}

\begin{figure}[h!]
\centering
\includegraphics[trim={0mm 0mm 0mm 0mm}, clip,width=0.97\linewidth]{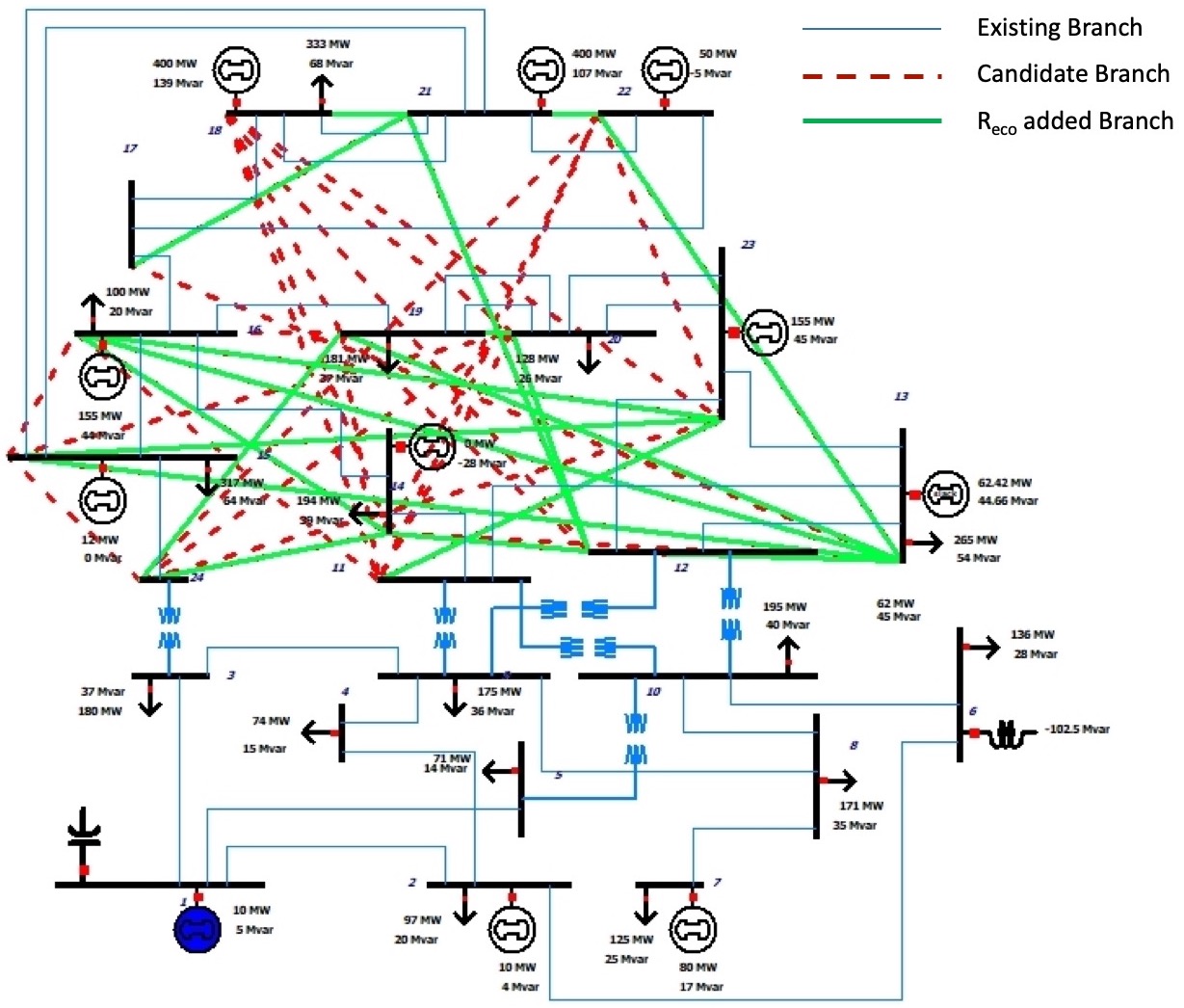}
  \caption{R\textsubscript{ECO}-oriented IEEE 24 Bus RTS network topology with 50 candidate branches (21 branches are constructed)} 
  \label{24case}
  
\end{figure}

The IEEE 24 Bus Reliability Test System (RTS) \cite{zimmerman2016matpower} has 24 buses and 37 branches. 
With 24 buses, there are 276 links that can be selected as candidate branches to expand the network structure. 
Fig \ref{24case} shows the R\textsubscript{ECO} optimized network for the IEEE 24 Bus RTS system with 50 candidate branches ($2^{50}$ different network structures), 
and 21 branches are added after the optimization. 

Table \ref{24 bus result} shows the results of all four scenarios for the IEEE 24 Bus RTS cases. The results of \textbf{Achieved R\textsubscript{ECO}} show that the optimized networks have a higher value of R\textsubscript{ECO} than the original case, and the \textit{-Str-OPF} networks have a higher value of R\textsubscript{ECO} than the \textit{-Structure} networks (except for the 100 candidate branch scenario). The value of optimized R\textsubscript{ECO} is close to the 
‘Window of Vitality’ (0.3469-0.3679), which is the unique range of R\textsubscript{ECO} for the resilient ecosystems \cite{Borrett2014}.

With more branches constructed, the system has fewer real power losses but more reactive power losses, and the apparent power losses (MVA) are increasing as shown in Table \ref{24 bus result} (except IEEE 24 Bus RTS-200-Structure and -Str-OPF). 
However, the extra losses from the new branches do not incur extra operational cost. When we compared the \textit{-Structure} cases to the original case, the operational cost is reduced. 
On the other hand, the operational cost of all \textit{-Str-OPF} cases increases with a slightly higher R\textsubscript{ECO} values (except IEEE 24 Bus RTS-100-Str-OPF). 
\textcolor{black}{With the optimized output vectors, $P_{gen_{i}}$ and $\theta_{i}$, the generators are also more equally contributing to the power supply for improving the R\textsubscript{ECO}. Some expensive generators are generating more power, while some cheaper generators are producing less.} It shows that the operational cost will not change much if \textcolor{black}{only} the network structure is more robust. \textcolor{black}{As mentioned in Section \ref{Algorithm1}, each set of candidate branches is unique. With the increasing numbers of candidate branches, the number of added branches does not increase. The added redundancy does not necessarily depend on the number of candidate branches. This confirms that R\textsubscript{ECO} can \textit{strategically} construct the network structure and operate power systems to improve the system's resilience and maintain power system constraints.%
}

\begin{table}[t!]
\caption{Results of R\textsubscript{ECO}-Oriented Power Network Design for IEEE 24 Bus RTS }
\vspace{-.1in}

\begin{center}
\begin{adjustbox}{width=0.48\textwidth}
{\renewcommand{\arraystretch}{1.1}
    \begin{tabular}{>{\centering}m{2.2cm}| >{\centering\arraybackslash}m{1.2cm}|>{\centering}m{1.3cm}|>{\centering\arraybackslash}m{1.3cm}|>{\centering\arraybackslash}m{1.5cm}|>{\centering\arraybackslash}m{1.7cm}|>{\centering\arraybackslash}m{1.4cm}} 
    \hline
    \hline
    \textbf{Use Case}  & \textbf{Achieved R\textsubscript{ECO}} & \textbf{Operational Cost (\$/hr)} &  \textbf{Number of Added Branches} &  \textbf{Real Power Losses (MW)} &  \textbf{Reactive Power Losses (MVar)} & \textbf{Computation Time (seconds)}\\ 
    \hline
IEEE 24 Bus RTS  & 0.3382 & 62263 & 0 & 51.22 & 650.27 & 0 \\\hline 
IEEE 24 Bus RTS-50-Structure & 0.3492 & 61061	& 21 & 29.91 & 789.69 &1.74 \\ \hline 
IEEE 24 Bus RTS-50-Str-OPF 	& 0.3496 & 78433	& 21 & 24.01 & 799.34 & 1.74 \\  \hline 
IEEE 24 Bus RTS-100-Structure 
& 0.3514*  & 60716	& 25 & 19.92 & 891.52 & 8.54 \\  \hline 
IEEE 24 Bus RTS-100-Str-OPF & 0.3502 & 71582	& 25 & 19.19 & 898.74 & 8.54  \\ \hline 
IEEE 24 Bus RTS-150-Structure  & 0.3454 & 60925	& 12 & 24.19 & 682.01 & 75.90 \\ \hline 
IEEE 24 Bus RTS-150-Str-OPF  & 0.3459 & 75525	& 12 & 22.39 & 691.87 & 75.90  \\ \hline 
IEEE 24 Bus RTS-200-Structure & 0.3474 & 61432	& 8 & 34.40 & 598.75 &23.02 \\ \hline 
IEEE 24 Bus RTS-200-Str-OPF & 0.3479 & 78211	& 8 & 27.82 & 617.91 & 23.02 \\
    \hline
    \hline
\end{tabular}}
\end{adjustbox}
\end{center}
\label{24 bus result}
\end{table}

\subsection{ACTIVSg200}

\begin{figure}[h!]
\centering
\includegraphics[trim={0mm 0mm 0mm 0mm}, clip,width=0.97\linewidth]{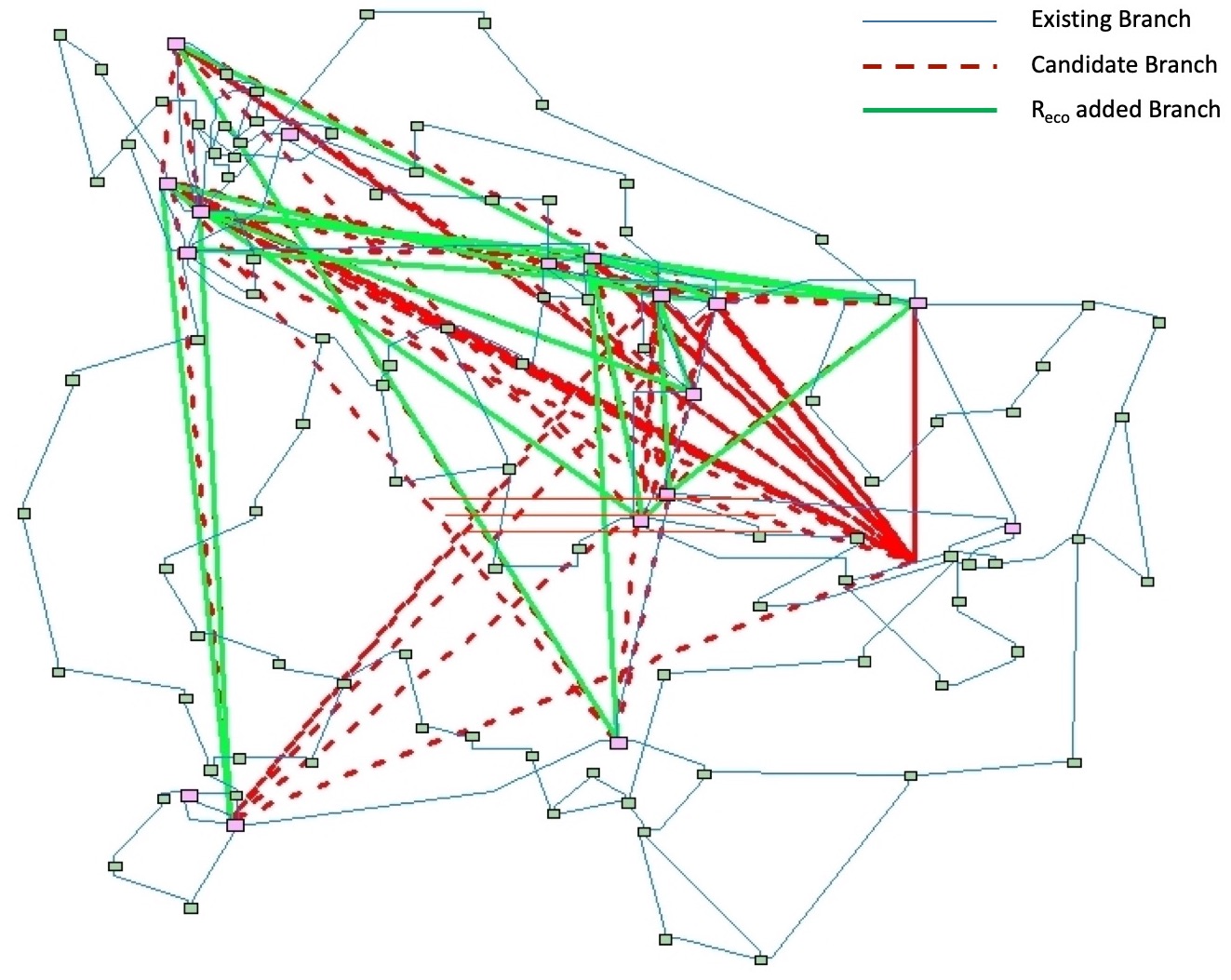}
  \caption{R\textsubscript{ECO}-oriented ACTVISg200 network topology with 50 candidate branches (26 branches are constructed)} 
  \label{200case}
  
\end{figure}

The ACTIVSg200 case \cite{synthetic} has 200 buses and 246 branches. 
With 200 buses, there are 19900 links that can be selected as candidate branches, which contains $2^{19900}$ different network structures to be explored. Fig \ref{200case} shows the R\textsubscript{ECO} optimized network for the ACTIVSg200 system with 50 candidate branches ($2^{50}$ different network structures), and 26 branches are added after the optimization. 

\begin{table}[t!]
\caption{Results of R\textsubscript{ECO}-Oriented Power Network Design for ACTIVSg200}
\vspace{-.1in}
\begin{center}
\begin{adjustbox}{width=0.48\textwidth}
{\renewcommand{\arraystretch}{1.1}
    \begin{tabular}{>{\centering}m{1.9cm}| >{\centering\arraybackslash}m{1.2cm}|>{\centering}m{1.3cm}|>{\centering\arraybackslash}m{1.3cm}|>{\centering\arraybackslash}m{1.5cm}|>{\centering\arraybackslash}m{1.7cm}|>{\centering\arraybackslash}m{1.4cm}} 
    \hline
    \hline
    \textbf{Use Case}  & \textbf{Achieved R\textsubscript{ECO}} & \textbf{Operational Cost (\$/hr)} &  \textbf{Number of Added Branches} &  \textbf{Real Power Losses (MW)} &  \textbf{Reactive Power Losses (MVar)} & \textbf{Computation Time (seconds)}\\ 
    \hline 
ACTIVSg200&   0.2510 & 49000 & 0 & 24.77 & 435.64 & 0 \\ \hline 
ACTIVSg200-50-Structure	 & 0.2651 & 48909	& 26 & 19.21 & 598.83 &58.64 \\ \hline 
ACTIVSg200-50-Str-OPF 	& 0.2655 & 49725	& 26 & 18.04 & 593.64 & 58.64 \\  \hline 
ACTIVSg200-100-Structure 	& 0.2578 & 51264	& 15 & 20.92 & 523.41 & 35.46\\  \hline 
ACTIVSg200-100-Str-OPF  & 0.2599 & 51264	& 15 & 21.55 & 526.00 & 35.46 \\ \hline 
ACTIVSg200-150-Structure & 0.2531 & 48923	& 5 & 22.23 & 471.84 & 84.09 \\ \hline 
ACTIVSg200-150-Str-OPF & 0.2557 & 50044	& 5 & 22.73 & 470.65 & 84.09 \\ \hline 
ACTIVSg200-200-Structure  & 0.2671 & 52104	& 51 & 17.63 & 748.15 & 45.80 \\ \hline 
ACTIVSg200-200-Str-OPF 
& 0.2708*  & 52014	& 51 & 16.75 & 764.93 & 45.80 \\
    \hline
    \hline
\end{tabular}}
\end{adjustbox}
\end{center}
\label{200 result}
\end{table}

All four scenarios are successfully solved and the results are shown in Table \ref{200 result}. Compared to the IEEE 24 Bus RTS, the \textbf{Achieved R\textsubscript{ECO}} values are much smaller in ACTIVSg200 cases. The original synthetic power grids are highly close to the real U.S power grids, which are quite sparse and efficient. Considering there are  $2^{19900}$ different structures that can be explored, the created candidate branches may not have the \textit{exact} optimal structure. 
Thus, the R\textsubscript{ECO} for this synthetic grid is not improved as much as the IEEE 24 Bus RTS system. 

For the ACTIVSg200 cases, all \textit{-Str-OPF} networks have higher R\textsubscript{ECO} than their corresponding \textit{-Structure} networks. The new built branches incur extra whole power losses. Similar to the IEEE 24 RTS case, the real power losses also decrease while the reactive power losses increase. The operational cost of the ACTIVSg200-50-Structure and ACTIVSg200-150-Structure cases are less than the original operational cost even though there are extra branches and losses. The operational cost of other R\textsubscript{ECO}-oriented cases increase slightly compared to the original case. 
The number of built branches does not increase with the increasing of candidate branches. This also demonstrates the R\textsubscript{ECO} is \textit{strategically} constructing the network structure and operating power systems.

\section{Network Analyses}\label{results}

The optimized networks are analyzed and compared with their original network for their reliability under \textit{multi-hazard} scenarios and network properties regarding their structure and power flow distribution. All analyses are performed using AC power flow model. 



\subsection{Network Reliability Analysis}\label{ReliabilityAnalysis}


The \textit{multi-hazard} contingencies are applied as different levels of \textit{N-x} contingencies for each case. For \textit{x=1}, they are planned contingencies; for \textit{x$>$1}, they are \textit{unexpected} contingencies. Under the contingencies, if there is one branch's power flow is over the limit or the voltage magnitude is out of the required limit, it is counted as \textbf{one} violation. If the power flow cannot be solved, then the contingency is marked as \textbf{unsolved}. 

\textcolor{black}{As for different case studies, the generation of N-\textit{x} contingencies are different since the IEEE 24 Bus RTS system is relatively small compared with the ACTIVSg200 system.} 
\textcolor{black}{For the IEEE 24 Bus RTS cases, comprehensive \textit{N-1}, \textit{N-2} and \textit{N-3} contingency analyses are performed for all power system components, including branch, bus and generator. 
The loss of any bus can cause more elements to be disconnected simultaneously. Thus, the \textit{N-3 bus} contingencies can cause multiple components (generators and branches) disconnected. This can have a similar impact on generator unavailability like the Texas Winter Storm \cite{busby2021cascading}. 
For the ACTIVSg200 cases, a comprehensive \textit{N-2} and \textit{N-3} contingency analysis is difficult to complete, due to the large number of components. The \textit{N-1} contingency analysis is done for the branch, bus, and substations, respectively. Since all generators in ACTIVSg200 case are connected through transformers, the \textit{N-1} branch contingencies include all \textit{N-1} generator contingencies. The loss of one bus and one substation can catastrophically impact the entire system with multiple components (\textit{N-x}) disconnected. It provides validation of the redesigned system's ability to tolerate disturbances and maintain functionality securely. For the ACTIVSg200 cases, the \textit{unexpected} critical \textit{multi-hazard} contingencies from \cite{narimani2021generalized,huang2021toward} are also considered. As mentioned in Section \ref{multihazard}, such critical \textit{N-x} contingencies (\textit{x} ranges from 3 to 8) are selected through LODFs and GBC as multiple branches widely spread in the system, whose loss may cause catastrophic impact to the system. Such critical contingencies are both geographically wide spread and statistically rare, which make them a touchstone to study
resilience in large-scale 
systems. }
All the contingency analyses investigated here are performed without remedial actions. The basic control mechanism, such as automatic generation control (AGC) and automatic voltage regulator (AVR), are retained at their original settings. This provides a fair study about each system's inherent ability to tolerate unexpected \textit{multi-hazard} disturbances and maintain functionality securely, thus justifying the improvement of resilience.

\begin{figure}[h]
\centering
\includegraphics[trim={1.2mm 2cm 1.2mm 2mm},scale=0.29]{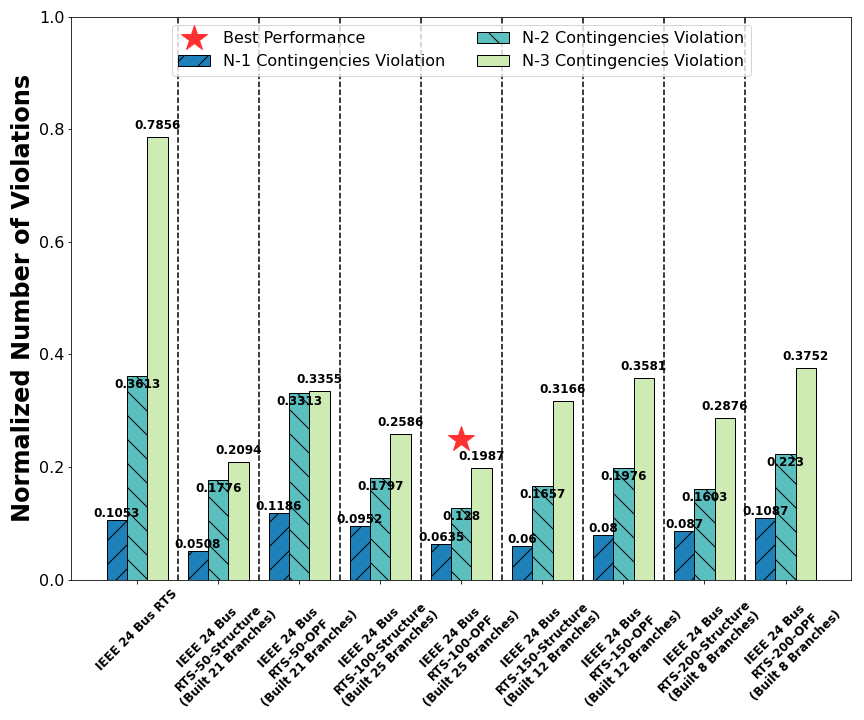}
  \label{24F}
 \caption{Normalized Violations Comparison of R\textsubscript{ECO}-Oriented Power Network for all variations of IEEE 24 Bus RTS cases (Table \ref{24 bus result})}\label{24_reliability}

\end{figure}

\begin{figure}
\centering
\includegraphics[trim={1.2mm 1.8cm 1.2mm 2mm},scale=0.29]{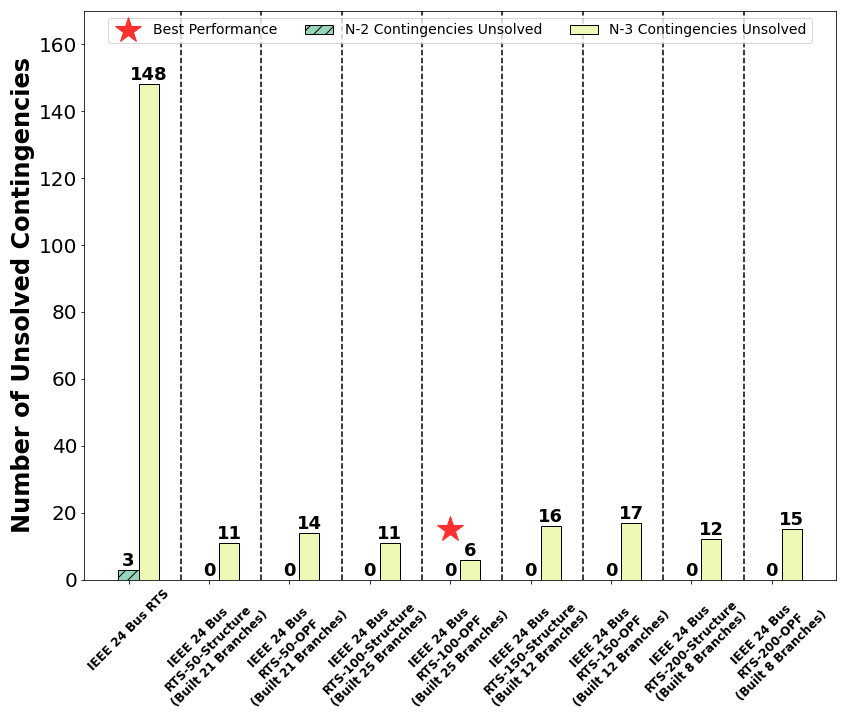}
  \label{24F}
 \caption{Unsolved Contingencies Comparison of R\textsubscript{ECO}-Oriented Power Network for all variations of IEEE 24 Bus RTS cases (Table \ref{24 bus result})}\label{24_reliability_unsolved}

\end{figure}

With more branches built after the optimization, there are more \textit{N-1}, \textit{N-2} and \textit{N-3} contingencies than the original case, especially for the IEEE 24 Bus RTS case. To fairly compare the reliability, we then normalize the number of violations with the total number of \textit{N-x} contingencies. Fig.  \ref{24_reliability} shows the normalized violations (total violations/total number of contingencies), and Fig. \ref{24_reliability_unsolved} shows the unsolved \textit{N-2} and \textit{N-3} contingencies for all variations of the IEEE 24 Bus RTS cases. Overall, the R\textsubscript{ECO}-oriented network structure and operation schemes are more reliable than the original case with far fewer normalized violations and unsolved contingencies. 
With the proposed R\textsubscript{ECO}-oriented approach, the \textit{unsolved} \textit{N-2} contingencies are completely resolved and the number of \textit{unsolved} \textit{N-3} contingencies reduced from 148 to less than 20. 
This ensures the observability of the system during disturbances and shows an outstanding improvement of resilience. The IEEE 24 Bus RTS-100-OPF case has the best performance among all cases. Even though its achieved R\textsubscript{ECO} is smaller than the corresponding \textit{-Structure} case, they have the same network structure. The redundant network structure contributes to the improved resilience.

\begin{figure}
\centering
\includegraphics[trim={1.2mm 1cm 1.2mm 1cm}, scale=0.19]{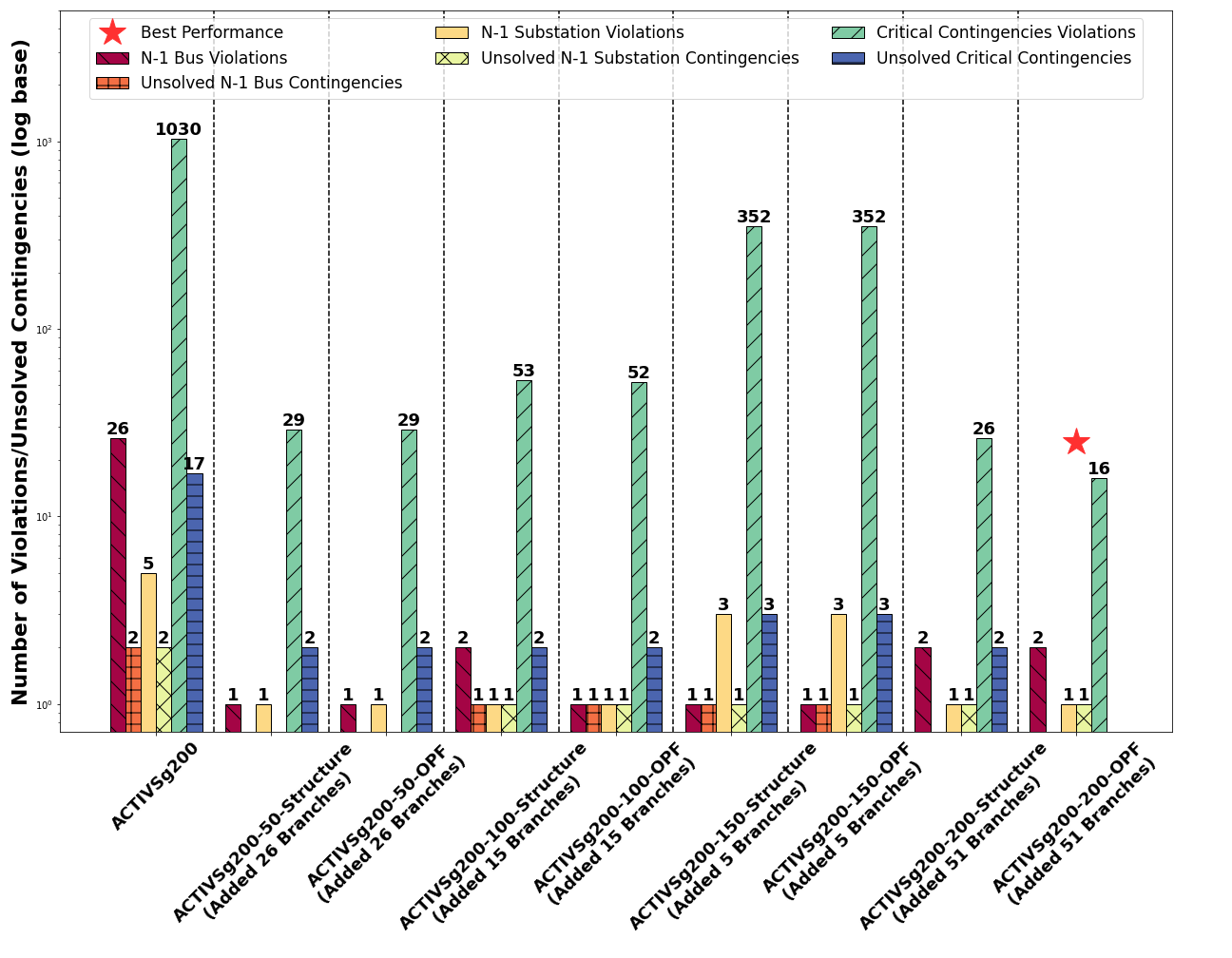}
 \caption{Reliability Comparison of R\textsubscript{ECO}-Oriented Power Network for all variations of ACTIVSg200 cases (Table \ref{200 result})}\label{200_reliability}

\end{figure}

Fig \ref{200_reliability} shows the contingency analysis for all ACTIVSg200 cases. Overall, the R\textsubscript{ECO}-oriented networks are much more resilient than the original network. All the optimized ACTIVSg200 cases maintain the same \textit{N-1 branch} reliability as the original network. Under the \textit{N-1} Bus, \textit{N-1} Substation contingencies, and \textit{unexpected} multi-hazard contingencies, all the R\textsubscript{ECO}-oriented networks are \textit{more} reliable with much fewer violations and unsolved situations. The ACTIVSg200-200-OPF has the best performance out of all the optimized networks with minimum violations and unsolved contingencies with the highest achieved R\textsubscript{ECO}. 

\subsection{Network Properties Analysis}

An entropy based network robustness metric (\textbf{$R_{CF}$}) is used to identify cascading failures in power systems \cite{kocc2013entropy}. The analysis of \textbf{$R_{CF}$} can capture how likely it is for the network to experience cascading failures. With a higher value of \textbf{$R_{CF}$}, the network is more robust and less likely to have a cascading failure \cite{kocc2013entropy}. The calculation of \textbf{$R_{CF}$} follows,

\begin{equation}
R_{CF} =\sum_{i=1}^{N}R_{n,i}\delta_{i}
\label{rcf}
\end{equation}
\begin{gather}
R_{n,i} =-\sum_{i=1}^{L}\alpha_{i}p_{i}log(p_{i}) \:\:\:  and \:\:\: \delta_{i} =\frac{P_{i}}{\sum_{j=1}^{N}P_{j}} 
\label{eq_tsupp}
\end{gather}

\noindent where $\alpha_{i}$ is the ratio between the maximum capacity and the load of corresponding line $i$; $p_i$ is the normalized flow values on the out-going links; 
$P_i$ is the total power distributed by node $i$ and $N$ is the number of nodes in the network.

All network structures are analyzed for typical complex network properties, including the average node degree (\textbf{$\bar{d}$}), clustering coefficient (\textbf{$\bar{c}$}), average betweenness centrality measures (\textbf{$\bar{b}$}) and average shortest path (\textbf{$\bar{l}$}) \cite{bollobas2013modern},

\begin{gather}
    \bar{d}= \frac{\sum e}{\sum n} \; ; \;\;\; \bar{c}=\frac{\sum_{i}\frac{\sum_{j,k}A_{ij}A_{jk}A_{ki}}{\sum_{j}A_{ij}(\sum_{j}A_{ij}-1)}}{\sum n} \; ; \;\;\;
    \\
     \bar{l}=\frac{\sum_{i,j} dist(v_{i},v_{j})}{\sum{i,j} has\_path(v_{i},v_{j})}  \; ; \;\;\;
     \bar{b}=\sum_{s,t\in V}\frac{\sigma(s,t|e)}{\sigma(s,t)}
    \label{network}
\end{gather}

\noindent where $e$ is the edge and $n$ is the node in graph; $\sum n$ is the total number of nodes in the graph; $A$ is the adjacency matrix of the graph; $\sigma(s,t)$ represents the number of shortest paths in the graph between $s$ and $t$; $\sigma(s,t|e)$ is the number of shortest paths in the graph between $s$ and $t$ that contain edge $e$.

The power flow distribution is also investigated by calculating the \textit{Mean} and \textit{Standard Deviation (STD)} of all branches' \textcolor{black}{real power flow (pf), reactive power flow (rf), and the line percentage of MVA limit (MVA\%) using Eqn. (\ref{mean}). For the power flow, the \textit{x\textsubscript{i}} are all branches' pf and rf, respectively. For the line percentage, the \textit{x\textsubscript{i}} are all branches' MVA\%.} The \textit{N} is the total number of branches. 

\begin{gather}
    \overline{x} =\frac{1}{N} \sum_{i=i}^{n} x_{i}\; ; \;\;\; s(x) = \sqrt{\frac{1}{N-1} \sum_{i=1}^N (x_i - \overline{x})^2}
    \label{mean}
\end{gather}

Table \ref{NA} shows the network properties for all network structures and the corresponding optimal power flow. The R\textsubscript{ECO}-oriented networks have better network properties than their original counterparts. All the R\textsubscript{ECO}-oriented networks have higher $R_{CF}$, showing that the R\textsubscript{ECO} corresponds to an improved $R_{CF}$ against cascading failures. Increasing $R_{CF}$ is found to highly correlate with increasing \textit{R\textsubscript{ECO}}, except the optimized results of the IEEE 24 Bus RTS with 100 candidate branches. \textcolor{black}{Although formulations of both R\textsubscript{ECO} and R\textsubscript{CF} are based on an entropy model, their modeling details are different. R\textsubscript{CF} is based on branch flow limits, while R\textsubscript{ECO} is based on network structure, flow magnitudes, and flow directions. There can be some discrepancies between these two metrics.}

All the R\textsubscript{ECO}-oriented networks have larger \textbf{$\bar{d}$} and \textbf{$\bar{c}$}, and reduced \textbf{$\bar{b}$} and \textbf{$\bar{l}$}. It shows these networks are more robust, reducing the significance of nodes (buses) and paths (branches) in the system, \textcolor{black}{which spreads out the system's risks, from both perspectives of severity and probability.}
For actual networks, the \textbf{$\bar{d}$} is in the range of (2.58, 2.61), the \textbf{$\bar{c}$} is in the range of (0.032, 0.058), the \textbf{$\bar{b}$} is in the range of (0.083, 0.40), and the \textbf{$\bar{l}$} is in the range of (14.2, 29.2) \cite{synthetic}. The results show that the optimized ACTIVSg200 networks' \textbf{$\bar{d}$} and \textbf{$\bar{c}$} are close to the actual systems, but the \textbf{$\bar{b}$} and \textbf{$\bar{l}$} are not. These can be explained by the way candidate branches were selected at their highest voltage level for each case, whose distance is shorter than branches between different voltage levels.

\begin{table*}[t!]
\caption{Network properties for all variations of the IEEE 24 Bus RTS and ACTIVSg200 systems.}
\vspace{-.1in}
\begin{center}
\begin{adjustbox}{width=0.98\textwidth}
 
\renewcommand\arraystretch{1.5}
\begin{threeparttable}
\begin{tabular}{c|c|c|c|c|c|c|c|c|c|c|c|c} 
    \hline
    \hline
    
    \textbf{Use Case}  & \textbf{Achieved \textit{R\textsubscript{ECO}}} & \textbf{$R_{CF}$}\cite{kocc2013entropy} & \textbf{$\bar{d}$} &  \textbf{$\bar{c}$} & \textbf{$\bar{b}$}& \textbf{$\bar{l}$} & Mean(pf) & STD(pf) & \textcolor{black}{Mean(rf)} & \textcolor{black}{STD(rf)} & \textcolor{black}{Mean(MVA\%)} & \textcolor{black}{STD(MVA\%)} \\ 
    \hline
    
IEEE 24 Bus RTS &    0.3382 & 1.121 & 2.833 & 0.03472 & 0.10063 & 3.2138 & 117.19	&86.83 & \textcolor{black}{27.95} & \textcolor{black}{23.52} 	&32.35	&19.04\\ \hline 

IEEE 24 Bus RTS-50-Structure &  0.3492  & 3.328 & 4 & 0.17698 & 0.07246 & 2.5942 &  67.01	&53.33 & \textcolor{black}{19.79} & \textcolor{black}{19.31} 	&19.54	&16.35 \\ \hline 

IEEE 24 Bus RTS-50-Str-OPF & 0.3496 & 3.413 & 4 & 0.17698 & 0.07246 & 2.5942 & 62.94	&51.00 & \textcolor{black}{21.46} & \textcolor{black}{19.86} 	&18.93	&16.11	 \\  \hline 

IEEE 24 Bus RTS-100-Structure & 0.3514 & 4.009 & 4.333 & 0.175 & 0.06637 & 2.4601 & 56.34	&43.71 & \textcolor{black}{18.91} & \textcolor{black}{19.02} 	&17.02	&15.97 \\  \hline 
IEEE 24 Bus RTS-100-Str-OPF \textcolor{red}{$\star$}  &  0.3502 &  4.043 & 4.333 & 0.175 & 0.06637 & 2.4601 & 51.47	&42.57 & \textcolor{black}{19.57} & \textcolor{black}{19.67} &16.48	&17.22   \\ \hline 

IEEE 24 Bus RTS-150-Structure  & 0.3454 & 2.758 & 3.5	& 0.09861 & 0.08037 & 2.7681 & 69.68	&60.11 & \textcolor{black}{20.32} & \textcolor{black}{22.60} 	&21.24	&18.22\\ \hline 

IEEE 24 Bus RTS-150-Str-OPF & 0.3459 & 2.495  & 3.5 & 0.09861 & 0.08037 & 2.7681 &68.54	&57.76 & \textcolor{black}{21.37} & \textcolor{black}{23.28} 	&21.13	&18.1   \\ \hline 

IEEE 24 Bus RTS-200-Structure &	0.3474  & 1.979 & 3.417 & 0.11389 & 0.07790 & 2.7138  & 97.94	&60.66 & \textcolor{black}{25.19} & \textcolor{black}{22.69} 	&26.94	&16.36 \\ \hline 

IEEE 24 Bus RTS-200-Str-OPF &	0.3479 & 2.094 & 3.417 & 0.11389 & 0.07790 & 2.7138  & 88.79	&52.59 & \textcolor{black}{24.74} & \textcolor{black}{24.22} 	&25.13	&15.72 \\
    
 \hline 
  \hline 
ACTIVSg200&    0.2510 & 1.565 & 2.46 & 0.03723 & 0.03531 & 7.9913 & 37.54	&56.65 & \textcolor{black}{7.70} & \textcolor{black}{9.95} 	&18.01	&19.06 \\ \hline 

ACTIVSg200-50-Structure	&  0.2651  & 2.785 & 2.65 & 0.04399 & 0.02899 &6.7396 & 34.00	&51.06 & \textcolor{black}{5.94} & \textcolor{black}{7.28} 	&16.11	&18.19  \\ \hline 

ACTIVSg200-50-Str-OPF &  0.2655 & 2.802 & 2.65 & 0.04399 & 0.02899 &6.7396 & 33.18	&50.61 & \textcolor{black}{5.96} & \textcolor{black}{7.27} 	&15.57	&17.30 	 \\  \hline 

ACTIVSg200-100-Structure & 	 0.2578 & 2.204 & 2.55 & 0.03751 & 0.03061 &7.0597 & 35.25	&53.02 & \textcolor{black}{6.52} & \textcolor{black}{8.31} 	&16.80	&18.41 \\  \hline 

ACTIVSg200-100-Str-OPF &  0.2599 &  2.168 & 2.55 & 0.03751 & 0.03061 &7.0597  & 35.54	&54.36 & \textcolor{black}{6.89} & \textcolor{black}{8.89} 	&16.24	&16.85  \\ \hline 

ACTIVSg200-150-Structure  &	 0.2531 & 1.706 & 2.51	& 0.03654 & 0.03145 &7.2272  & 36.90	&55.29 & \textcolor{black}{7.18} & \textcolor{black}{8.95} 	&17.59	&18.78 \\ \hline 

ACTIVSg200-150-Str-OPF & 0.2557 & 1.704 & 2.51	& 0.03654 & 0.03145 &7.2272 & 37.48	&58.73 & \textcolor{black}{7.51} & \textcolor{black}{9.40} 	&16.91	&16.94    \\ \hline 

ACTIVSg200-200-Structure   &	  0.2671  & 4.207 & 2.82 & 0.05346 & 0.02787 & 6.5181 & 30.50	&48.63 & \textcolor{black}{5.10} & \textcolor{black}{6.55} 	&14.61	&17.89  \\ \hline 

ACTIVSg200-200-Str-OPF \textcolor{red}{$\star$}  &	0.2708 & 4.207 & 2.82 & 0.05346 & 0.02787 & 6.5181  & 30.50	&48.63 & \textcolor{black}{5.22} & \textcolor{black}{6.69} 	&14.02	&16.36\\
 \hline  

    \hline
\end{tabular}

\begin{tablenotes}
     \item[1] \textcolor{red}{$\star$}: Best reliability.
  \end{tablenotes}
  
 \end{threeparttable}
\end{adjustbox}
\vspace{-.2in}
\end{center}

\label{NA}
\end{table*}

\textcolor{black}{The \textit{Mean} and \textit{STD} of all the branches' real power flow (pf), reactive power flow (rf), and line percentage of MVA limits (MVA\%) show that the R\textsubscript{ECO}-oriented networks distribute power flow more equally than the original network with reduced values in those measures. The Mean (pf) and STD (pf) in each \textit{-Str-OPF} network are smaller than the \textit{-Structure} network showing a more equally distributed real power flows,
while the \textit{-Structure} networks
more equally distribute reactive power flows than \textit{-Str-OPF} networks with smaller value of Mean (rf) and STD (rf). These facts could explain that even though the real power flows of the \textit{IEEE 24 Bus RTS-150-Str-OPF} are more equally distributed than \textit{IEEE 24 Bus RTS-150-Structure}, its R\textsubscript{CF} is smaller. The $\alpha_{i}$ for R\textsubscript{CF} (Eqn. \ref{eq_tsupp}) is the ratio between maximum line capacity considering real and reactive power, while the line loading in its calculation is only real power. With less equally distributed reactive power, its R\textsubscript{CF} can be reduced. 
Similarly, since [\textbf{T}] and R\textsubscript{ECO} only consider the real power flows, the reactive power flows can be distributed less equally to support the new built branches. Thus, the \textit{-Str-OPF} cases may less equally distribute the power flows regarding the loading capacity, with higher value of the Mean (MVA\%) and STD (MVA\%) than the corresponding \textit{-Structure} cases. 
In the optimized network, the reduced STD (pf), STD (rf), and STD (MVA\%), compared to their original distributions, show that the power flows are closer to each other, and the newly built branches do not cause power flow increases on other branches. This also shows the proposed approach does not cause Braess paradox.} 

\section{Discussion}
\label{discussion}

The proposed \textit{R\textsubscript{ECO} oriented approach for resilient power networks} is a typical NP-hard problem. Although the cases are different, the total number of topologies that the proposed approach explored is the same, which are $2^{50}$, $2^{100}$, $2^{150}$, and $2^{200}$. With the case size increasing, the computation time increases from 1.7 seconds to 84.09 seconds because of more power system variables ($P_{i}$ and $\theta_{i}$) and more complicated network structures. 
Thus, the computation time and complexity of the proposed approach depend on the number of power system variables and network structure.

Unlike a traditional network expansion problem using the AC power flow model \cite{bent2011transmission, taylor2012conic, mehrtash2021new}, this paper does not consider auxiliary equipment for new branches in the formulation.
The proposed R\textsubscript{ECO}-oriented power network design problem is based on the DC power flow model. The optimized network's reliability and network properties are then analyzed through solving the AC power flow model. From the analyses, the optimized power network structures with more equally distributed power flows have a greatly improved inherent ability to tolerate disturbances and maintain functionality securely. The improved resilience is shown by fewer operational violations and unsolved contingencies under the conventional \textit{N-1} and unexpected \textit{multi-hazard} contingencies. \textcolor{black}{The candidate branches are created in Algorithm \ref{create} without construction cost data. Thus, we are not able to perform as detailed a cost effectiveness analysis as in \cite{ma2019resilience, lagos2019identifying,shahidehpour2021tri, garifi2021transmission}.}

\begin{figure}[h!]
\centering
\includegraphics[trim={1.2mm 1.2mm 1.2mm 1.2mm}, clip,width=0.95\linewidth]{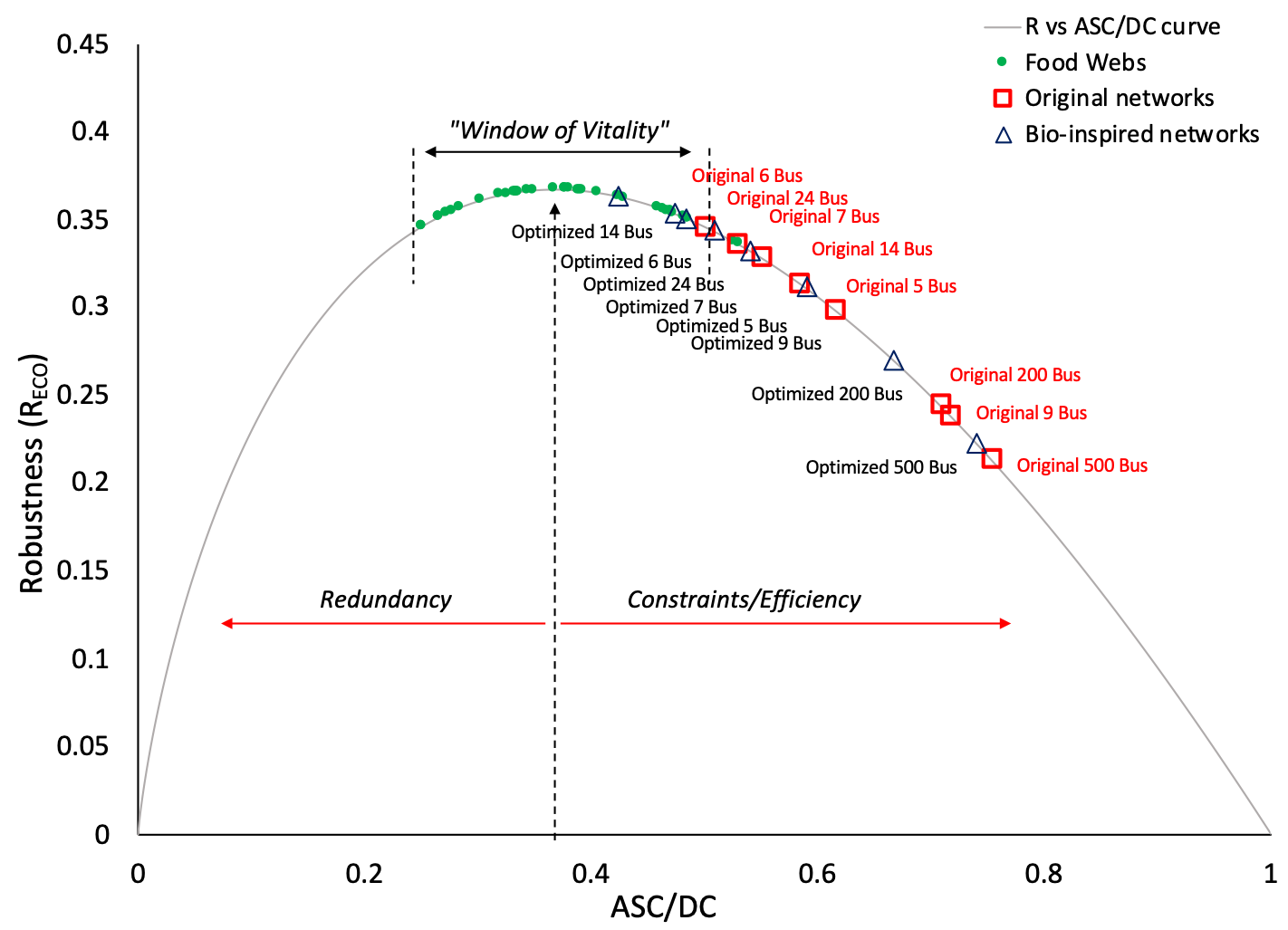}
  \caption{\textcolor{black}{R\textsubscript{ECO} curve for eight power grids and their R\textsubscript{ECO}-oriented versions, as well as a set of 38 food webs (Data source: \cite{panyam2019bioA}).}} 
  \label{rcurve}
  
\end{figure}

\textcolor{black}{Fig \ref{rcurve} shows the comparison of R\textsubscript{ECO} for eight power grids and a set of 38 food webs. The smaller power grids (5- to 14-bus cases) are optimized by a heuristic method in \cite{panyam2019bioA} and the larger power grids are optimized by the proposed approach in this paper. The R\textsubscript{ECO} of the food webs fall into the range of ‘Window of Vitality’, while the R\textsubscript{ECO} of the original power grids fall outside this range, especially the large and sparsely-connected power grids.
After the network optimization, their R\textsubscript{ECO} is improved, as well as their inherent ability to absorb disturbances. However, the R\textsubscript{ECO} is not 
within the ‘Window of Vitality’ for the cases in this paper. Two possible reasons for this are: (1) the desired ‘Window of Vitality’ values may be different for power systems compared to food webs,
and (2) the sets of candidate branches do not include \textit{all} network structures,
so
it is possible that the solution is not the \textit{exact} optimal structure recognized by R\textsubscript{ECO}.} Compared to the heuristic method in \cite{panyam2019bioA} \textcolor{black}{whose optimized cases are within ‘Window of Vitality,’} 
the proposed approach in this paper is more realistic with far fewer branches built. The approach in \cite{panyam2019bioA} is limited to a 14-bus system, thus we cannot directly compare both methods. 
The 14-bus case constructs 60 branches in \cite{panyam2019bioA} with a global heuristic search, while the proposed approach builds 51 branches for the ACTIVSg200 case.
It shows that the proposed R\textsubscript{ECO}-oriented approach with power flow constraints and limited search domain can realistically and strategically guide the power network design. Although the added branches slightly increase the operational cost for some scenarios, the improvement of reliability under different levels of \textit{N-x} contingencies and their network properties justifies this increased cost. \textcolor{black}{In \cite{huangreco}, R\textsubscript{ECO} was used to optimize the power flow distribution.
This paper uses R\textsubscript{ECO} to guide the power network design to further enhance its inherent capability to tolerate disturbances and maintain functionality securely. By strategically adding branches, the R\textsubscript{ECO}-oriented power networks are more resilient and survivable against multi-hazard contingencies, with much fewer violations and unsolved contingencies. From Table \ref{24 bus result}, Figure \ref{24_reliability} and \ref{24_reliability_unsolved}, 
the optimal R\textsubscript{ECO}-oriented IEEE 24 Bus RTS system reduces 70\% violations and 96\% unsolved contingencies with 25 added branches. 
From Table \ref{200 result} and Figure \ref{200_reliability} 
the optimal R\textsubscript{ECO}-oriented ACTIVSg200 case reduces 98\% violations and unsolved contingencies with 51 added branches. This level of resilience enhancement was not achieved in \cite{huangreco}.} \textcolor{black}{It shows that R\textsubscript{ECO} can be an \textit{accepted and unified} metric that captures power networks' inherent property of resilience.} 

The correlation among R\textsubscript{ECO}, \textit{R\textsubscript{CF}}, complex network properties and power flow distribution shows that the R\textsubscript{ECO}-oriented power network structure is more resilient against \textit{multi-hazard} and cascading failures due to the redundant network structure with equally distributed power flows. It is worth noting that the reactive power losses are predominant in transmission network as observed in Table \ref{24 bus result} and \ref{200 result}. With more branches built, the optimized systems have more reactive power losses. There should be some auxiliary equipment along with the new branches for reactive power compensation as in \cite{mehrtash2021new}. However, to investigate the influence of network structure to resilience, all systems keep their original real and reactive power capacity. Thus, the improvement of resilience solely comes from the R\textsubscript{ECO}-oriented network structure. With extra auxiliary devices for reactive power support, the optimized systems can be more reliable and resilient under the contingencies.
All above analyses demonstrate the effectiveness of using R\textsubscript{ECO} as a \textit{guidance} to \textit{strategically} design and operate power grids to improve its ability to absorb sudden and big disturbances in the system while maintaining their functions securely, thereby enhancing their resilience. 

\section{Conclusion}\label{sec:conclusion}

This work addresses a power system's need to withstand distributed threats arising from natural, accidental, and intentional causes that can create multi-hazard scenarios of $x$ elements across a wide area with severe impact. 
To achieve this, a power system resilient design approach is presented, inspired from long-term resilient ecosystems.  The resilience-oriented power grid network design problem is formulated and solved, with the goal to improve power systems’ inherent ability to tolerate disturbances and maintain functionality securely. The R\textsubscript{ECO}-oriented power networks are analyzed under \textit{N-x} contingencies, network properties, and operational cost. Results show the R\textsubscript{ECO}-oriented networks have fewer operational violations and unsolved contingencies with more redundant network structure and more equally distributed power flows. The R\textsubscript{ECO}-oriented optimization is generalizable as a resilient network design approach that improves a network’s ability to withstand unknown threats. 

\textcolor{black}{Future work can 
extend
upon this methodology from the following two aspects. On the one hand,
the impact of reactive power for the calculation and optimization of R\textsubscript{ECO} in power networks can be investigated for reactive power planning for better resiliency.
On the other hand, the economic factors, such as construction fee, electricity price, and penalty of unserved load, can be integrated with the proposed model to better understand the trade-offs between inherent resiliency and economics. 
Further, 
the projection of load growth and renewable energies integration can be taken into consideration for \textit{future} resilient and economic power network design.}

\section*{Acknowledgment}
The authors would like to acknowledge the Texas A\&M Energy Institute, the National Science Foundation under awards 1916142 and
2220347,
and the US Department of Energy Cybersecurity for Energy Delivery Systems program under award DE-OE0000895 for their support of this work.  

\bibliographystyle{IEEEtran}
\bibliography{hreference.bib}

\begin{IEEEbiography}[{\includegraphics[width=1in,height=1.25in,clip,keepaspectratio]{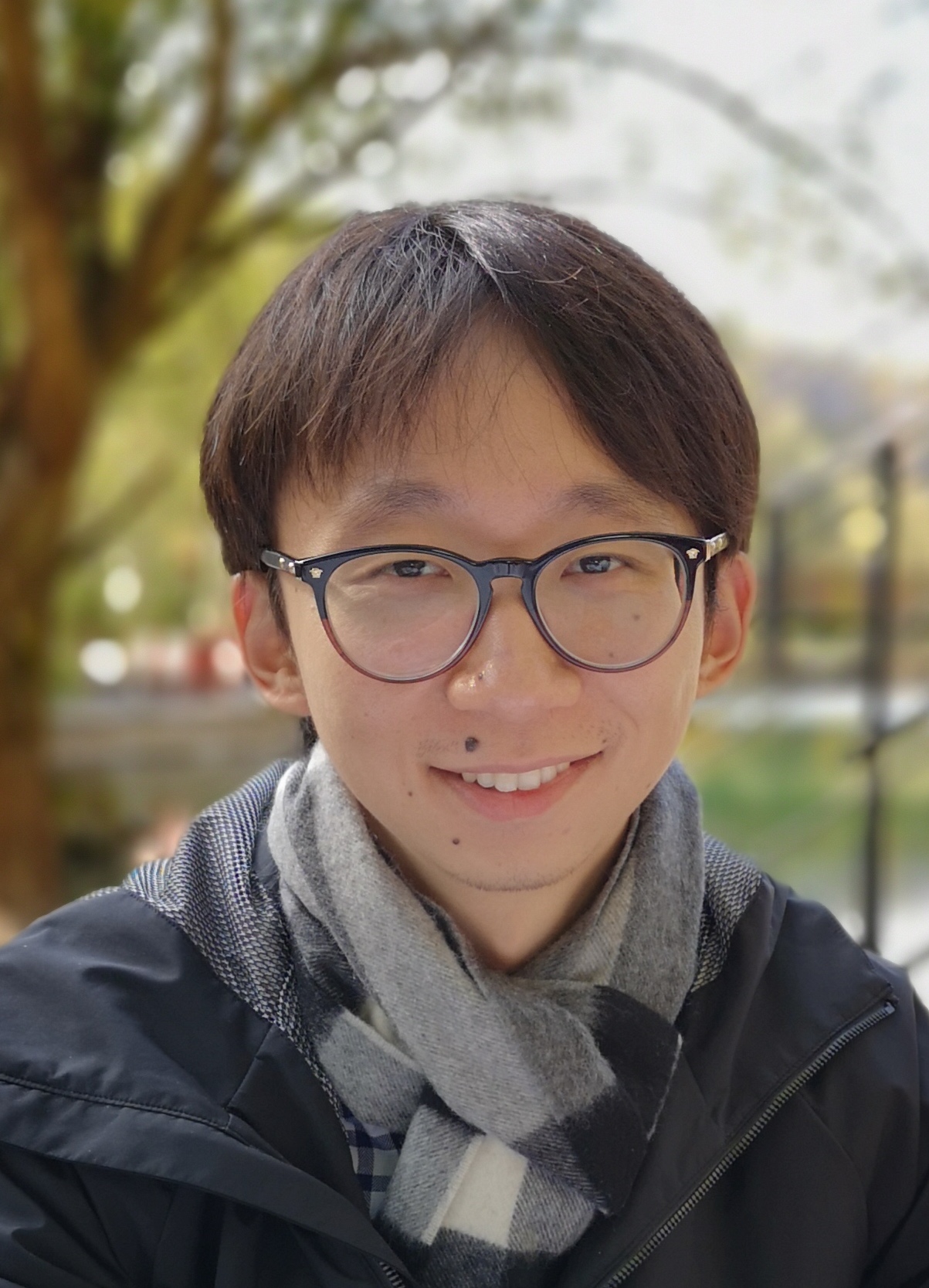}}]{Hao Huang (Member, IEEE)} received the B.S. degree in Electrical Engineering (Power System and Its Automation) from Harbin Institute of Technology, Harbin, Heilongjiang Province, China, in 2014, the M.S. degree in Electrical Engineering (Electric Power) from University of Southern California, Los Angeles, CA, USA, in 2016, and the Ph.D. degree in Electrical Engineering at Texas A\&M University, College State, TX, USA, in 2022. He is currently a Postdoctoral Research Associate at Princeton University. His research focuses on power system resilience, power system planning and operation, cyber-physical security, and scientific machine learning.
\end{IEEEbiography}

\begin{IEEEbiography}[{\includegraphics[width=1in,height=1.25in,clip,keepaspectratio]{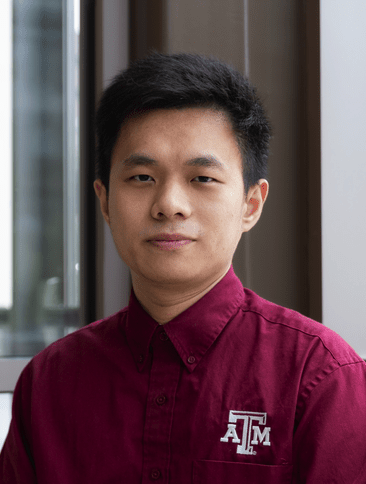}}]{Zeyu Mao (Member, IEEE)}
received the B.S. degree in Electrical Engineering from Chongqing University, Chongqing, China, in 2015, the M.S. degree in Electrical and Computer Engineering from the University of Illinois at Urbana–Champaign, IL, USA, in 2017 and the Ph.D. degree in Electrical Engineering at Texas A\&M University, College State, TX, USA, in 2022. He is currently a Quantitative Trading Analyst at DRW Holdings. His research interests include power system market, power system cyber-physical modeling, data-driven power system control, and sparse matrix ordering.
\end{IEEEbiography}

\begin{IEEEbiography}[{\includegraphics[width=1in,height=1.25in,clip,keepaspectratio]{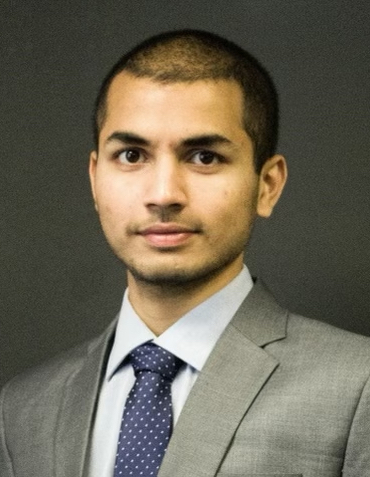}}]{Varuneswara Panyam} received the B.S. degree in Mechanical Engineering from Shiv Nadar University, Greater Noida, Uttar Pradesh, India, in 2016, and the M.S. degree in Mechanical Engineering from Texas A\&M University, College State, TX, USA, in 2018. He is currently working at Rani Therapeutics.
\end{IEEEbiography}

\begin{IEEEbiography}[{\includegraphics[width=1in,height=1.25in,clip,keepaspectratio]{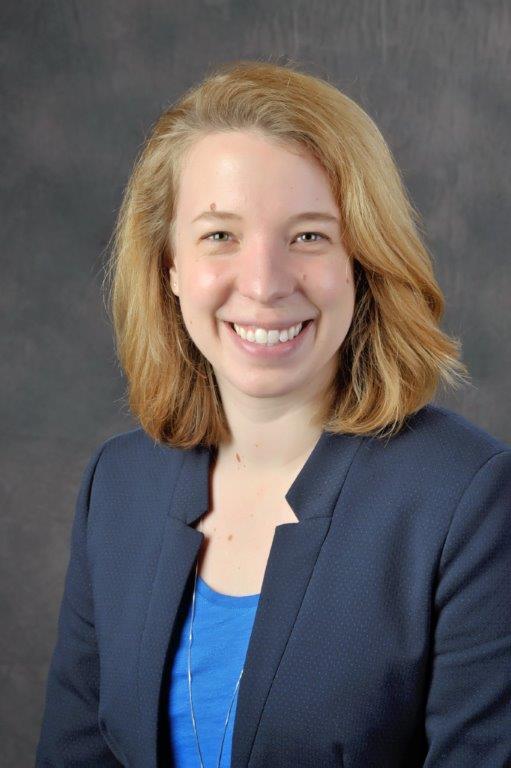}}]{Astrid Layton} received the B.S. degree in Mechanical Engineering from the University of Pittsburgh in Pittsburgh, PA in 2009 and the Ph.D. degree in Mechanical Engineering from Georgia Institute of Technology in Atlanta, Georgia in 2014, respectively. She is currently an Assistant Professor at Texas A\&M University in the Mechanical Engineering department. Her research looks at bio-inspired network design problems, focusing on the use of biological ecosystems as inspiration for the design of sustainable and resilient complex human networks and systems.
\end{IEEEbiography}


\begin{IEEEbiography}[{\includegraphics[width=1in,height=1.25in,clip,keepaspectratio]{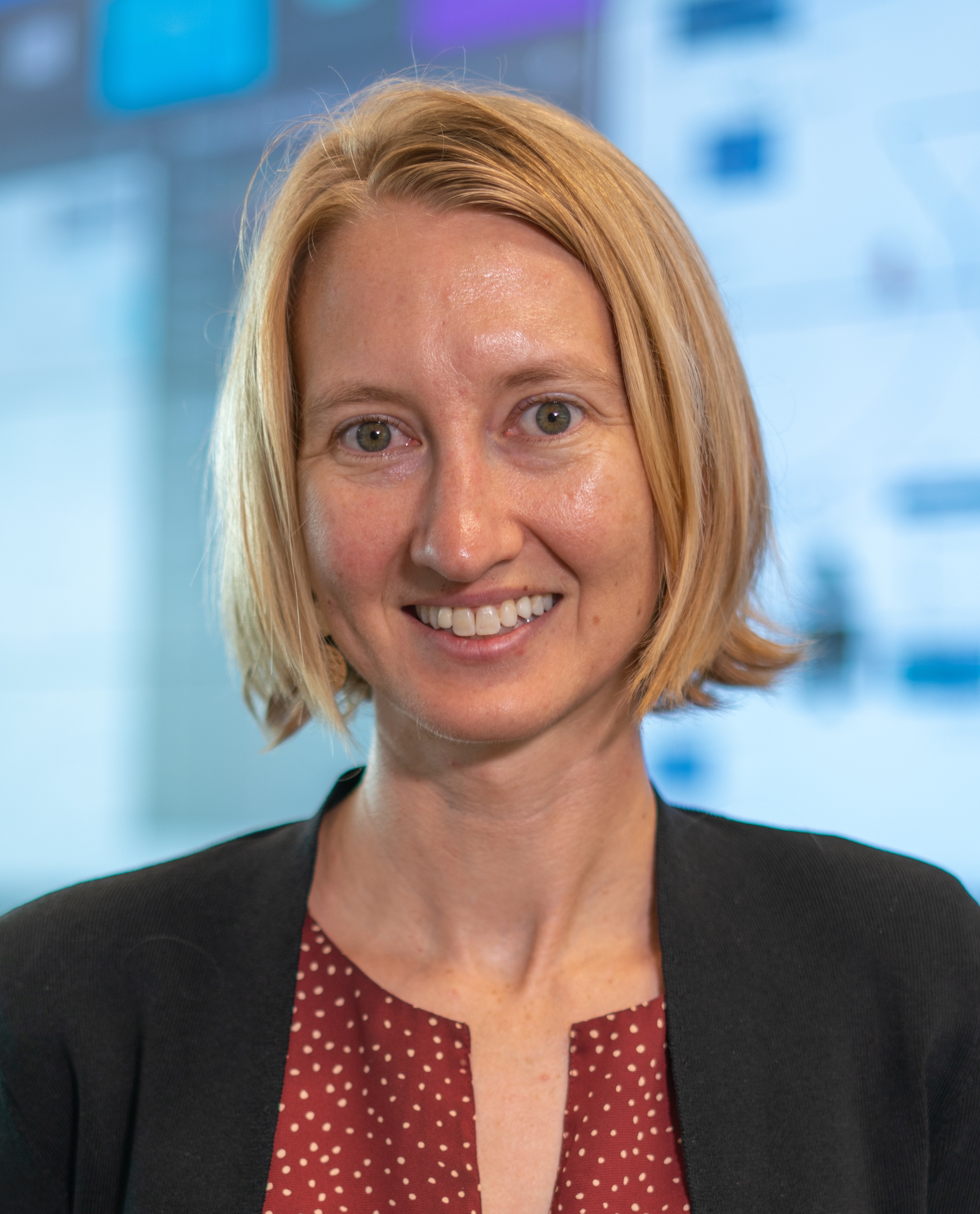}}]{Katherine R. Davis (Senior Member, IEEE)} received the B.S. degree from The University of Texas at Austin, Austin, TX, USA, in 2007, and the M.S. and Ph.D. degrees from the University of Illinois at Urbana–Champaign, Champaign, IL,
USA, in 2009 and 2011, respectively, all in electrical engineering. She is currently an Assistant Professor of electrical and computer engineering with Texas A\&M University. Her research interests include Operation and Control of Power Systems, Interactions between Computer Networks and Power Networks, Security-oriented Cyber-physical Analysis Techniques, Data-driven and Model-based Coupled Infrastructure Analysis and Simulation.
\end{IEEEbiography}

\end{document}